\documentclass[conference]{IEEEtran}
\IEEEoverridecommandlockouts
\usepackage{cite}
\usepackage{amsmath,amssymb,amsfonts}
\usepackage{graphicx}
\usepackage{textcomp}
\usepackage[bookmarks=false]{hyperref}
\usepackage{caption,amsmath,graphicx,url,amssymb,latexsym,epsfig,tabularx,setspace,multirow,threeparttable,longtable,pdflscape,tabu,comment,subfigure,textcomp,xparse}
\usepackage[noend]{algpseudocode}
\usepackage{algorithm}
\usepackage{color,soul}
\usepackage{forest, tikz}
\usepackage{xspace}
 \usepackage{hhline} 
\usepackage{enumitem}
\usepackage{makecell}

\usepackage{todonotes}

\def\selfjoinglobal{\textsc{GPUSelfJoinGlobal}\xspace}
\def\unicomp{\textsc{unicomp}\xspace}

\def\rtree{\textsc{CPU-RTree}\xspace}
\def\gpu{\textsc{GPU-SJ}\xspace}
\def\ego{\textsc{SuperEGO}\xspace}

\def\datasetsynParen{(\textit{Syn})\xspace}
\def\datasetsyn{\textit{Syn-}\xspace}
\def\dsynthaa{\textit{Syn2D2M}\xspace}
\def\dsynthab{\textit{Syn3D2M}\xspace} 
\def\dsynthac{\textit{Syn4D2M}\xspace} 
\def\dsynthad{\textit{Syn5D2M}\xspace} 
\def\dsynthae{\textit{Syn6D2M}\xspace}

\def\dsynthba{\textit{Syn2D10M}\xspace}
\def\dsynthbb{\textit{Syn3D10M}\xspace}
\def\dsynthbc{\textit{Syn4D10M}\xspace} 
\def\dsynthbd{\textit{Syn5D10M}\xspace} 
\def\dsynthbe{\textit{Syn6D10M}\xspace}

\def\datasetgeo{\textit{SW-}\xspace}
\def\dgeoaa{\textit{SW2DA}\xspace} 
\def\dgeoad{\textit{SW2DB}\xspace}
\def\dgeoba{\textit{SW3DA}\xspace} 
\def\dgeobd{\textit{SW3DB}\xspace}

\def\datasetsdss{\textit{SDSS-}\xspace}
\def\sdssa{\textit{SDSS2DA}\xspace} 
\def\sdssc{\textit{SDSS2DB}\xspace}

\begin{document}

\title{GPU Accelerated Self-join for the Distance Similarity Metric}

\author{\IEEEauthorblockN{Michael Gowanlock}
\IEEEauthorblockA{\textit{School of Informatics, Computing, \& Cyber Systems} \\
\textit{Northern Arizona University}\\
Flagstaff, AZ, 86011\\
michael.gowanlock@nau.edu}
\and
\IEEEauthorblockN{Ben Karsin}
\IEEEauthorblockA{\textit{Department of Information and Computer Sciences} \\
\textit{University of Hawaii at Manoa}\\
Honolulu, HI, 96822\\
karsin@hawaii.edu}
}

\maketitle

\begin{abstract}
The self-join finds all objects in a dataset within a
threshold of each other defined by a similarity metric. As such, the self-join is a building block for the field of databases and data mining, and is employed in Big Data applications. 
In this paper, we advance a GPU-efficient algorithm for the similarity self-join that uses the
Euclidean distance metric. The search-and-refine strategy is an efficient approach for low dimensionality datasets, as index searches degrade with increasing dimension (i.e., the curse of dimensionality). Thus, we target the low dimensionality problem, and compare our GPU self-join to a search-and-refine implementation, and a state-of-the-art parallel algorithm. 
In low dimensionality, there are several unique challenges associated with efficiently solving the self-join problem on the GPU.   Low dimensional data often results in higher data densities, causing a significant number of distance
calculations and a large result set.  As dimensionality increases,
 index searches become increasingly exhaustive, forming a
performance bottleneck. We advance
several techniques to overcome these challenges
using the GPU. The techniques we propose include a
GPU-efficient index that employs a bounded search, a batching scheme to accommodate large
result set sizes, and a reduction
in distance calculations through duplicate search removal. Our GPU self-join outperforms both search-and-refine and state-of-the-art algorithms.
\end{abstract}


\section{Introduction}\label{sec:intro}
The similarity self-join is a routine operation, and is described as follows: given a dataset of objects, find all objects that have common attributes based on a similarity metric. In spatial applications, the problem typically focuses on distance metrics to find points that are near each other. However, in principle, distance similarity  is a common metric, and can be applied to non-spatial applications as well. We focus on the distance similarity self-join that finds all points that are within a distance $\epsilon$ of each other using Euclidean distance. 

Similarity self-joins are building blocks of existing data analysis algorithms. For example, the DBSCAN clustering algorithm requires range queries that search the neighborhood of all data points to find those within a given distance~\cite{ester1996density}.  Other algorithms require range queries that constitute similarity joins, such as mining spatial association rules~\cite{koperski1995discovery}, information retrieval optimization~\cite{Bayardo:2007:SUP:1242572.1242591}, the OPTICS algorithm~\cite{Ankerst:1999:OOP:304182.304187}, and time series data analysis~\cite{agrawal1993efficient}. Therefore, the self-join is a fundamental component of established data analysis methods~\cite{Bohm:2000:HPC:354756.354832,bohm2009} and, as such, is used in many Big Data applications.

Self-joins and the related similarity join typically target either low or high dimensionality in the literature. This is because the methods used for the self-join in low dimensionality are often unsuitable for high dimensional data, and vice versa. As described in~\cite{kalashnikov2013}, a typical approach in low-dimensionality is to use the search-and-refine strategy as follows: \emph{search} an index for points that may be within the search radius of a query point, which generates a candidate set, and then \emph{refine} these points by performing the distance calculation between the query point and all points in the candidate set. This technique has been shown to be very efficient for low dimensional data; however, index performance degrades with dimensionality.  
We consider data up to 6-D, which is within the regime where index search performance degradation is not prohibitive for canonical spatial indexing schemes, and thus allows us to make a direct comparison to the search-and-refine approach.

\begin{figure}[th]
\centering
\subfigure[]{
        \includegraphics[width=0.22\textwidth, trim={0.5cm 0 0.5cm 0}]{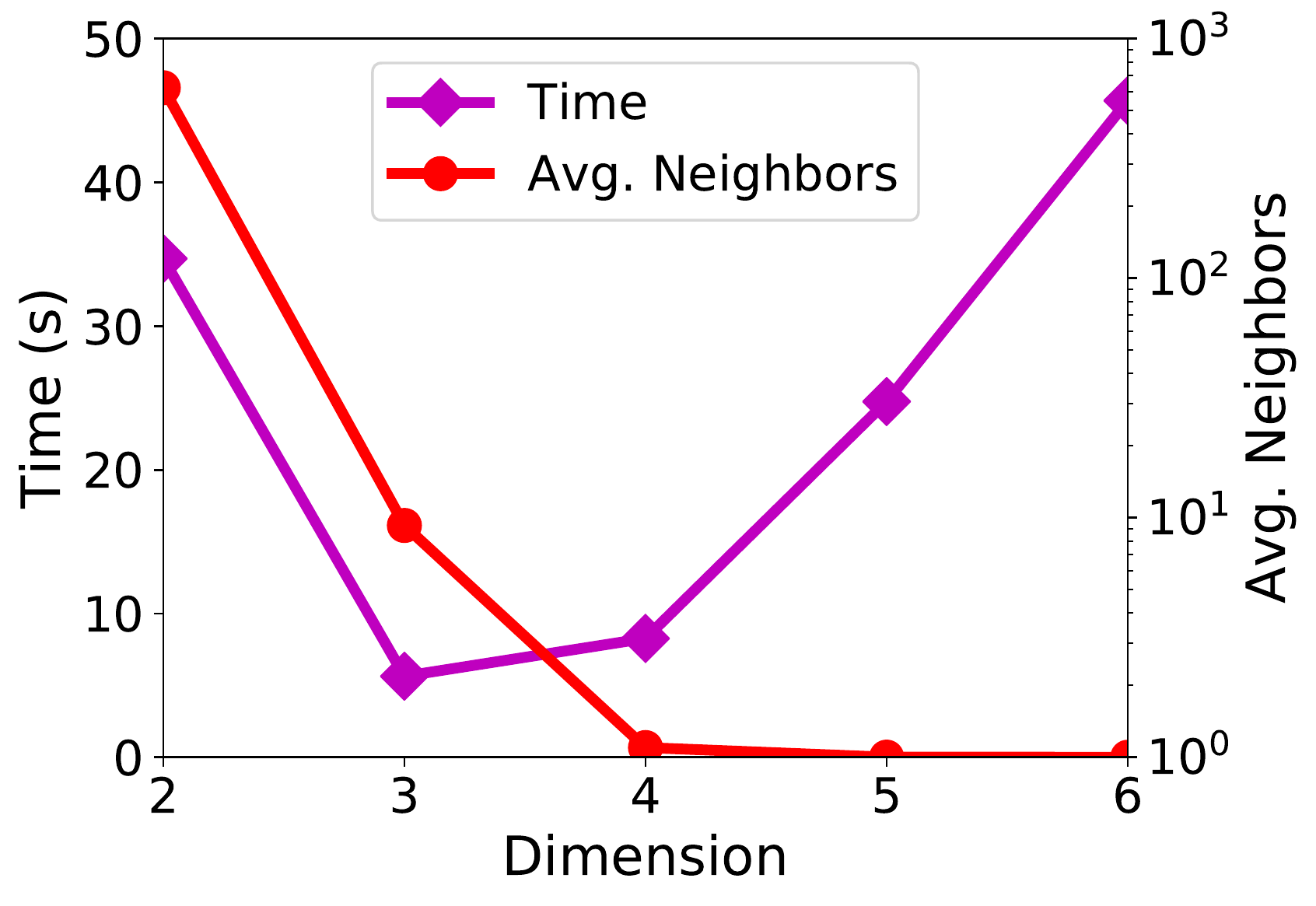}
    }
\subfigure[]{
    \includegraphics[width=0.22\textwidth, trim={0.5cm 0 0.5cm 0}]{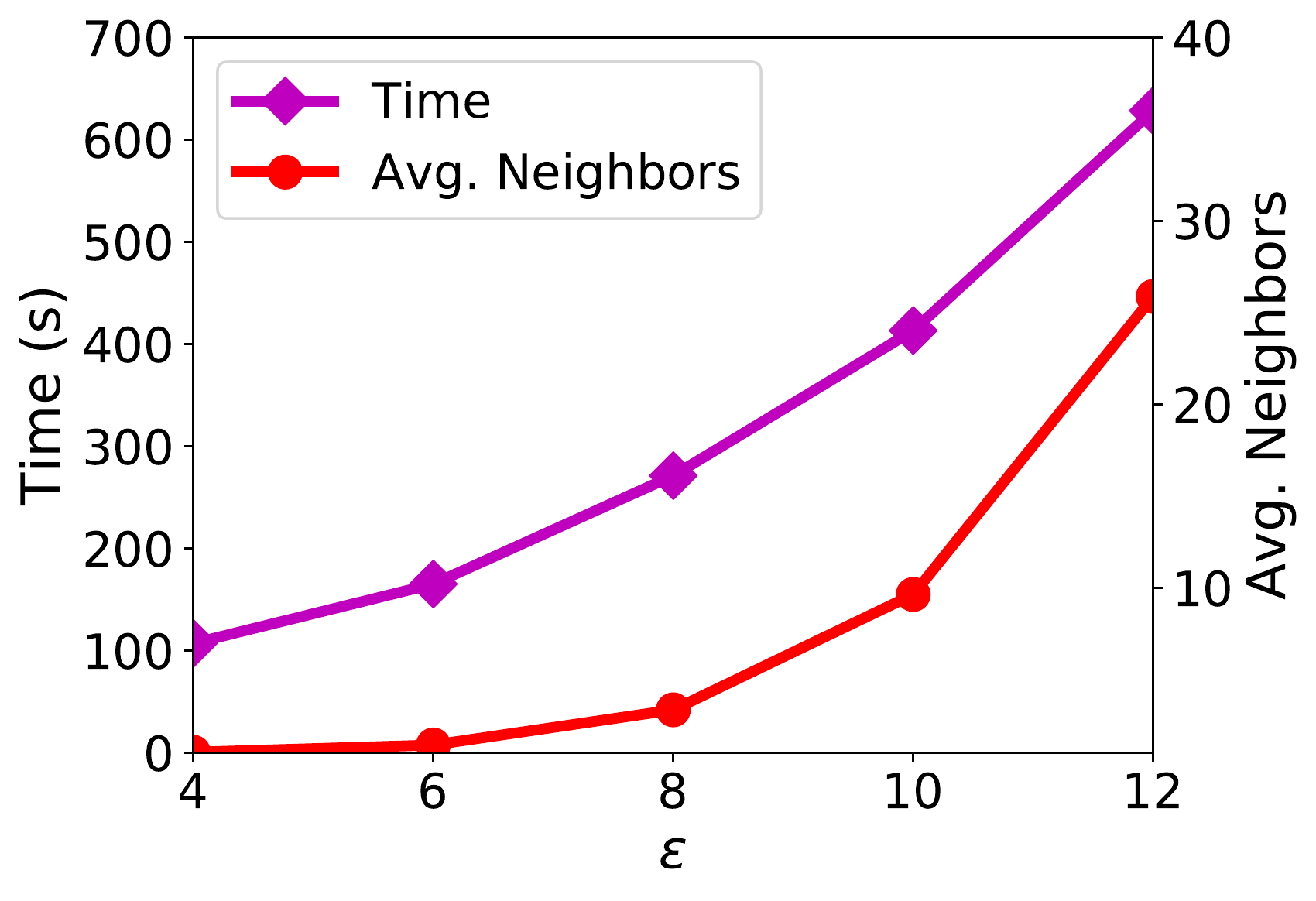}
    }
    \caption{Problem overview: (a) Computing the distance similarity self-join on 2 million data points in 2--6 dimensions with the distance $\epsilon=1$ using an R-tree index, where each dimension is of equal length and the points are uniformly distributed in a (hyper)cube. (b) Time vs. $\epsilon$ for the 6-D dataset shown in (a). Datasets (\datasetsyn) are described in Section~\ref{sec:datasets}.}
   \label{fig:motivation}
\end{figure}

We elaborate on performance as a function of dimensionality. Figure~\ref{fig:motivation}~(a) plots the self-join response time and the average number of neighbors per point vs. dimension for datasets with 2 million points that are indexed in the R-tree~\cite{Guttman-R_tree} on the CPU. Since the number of data points are kept constant, as the dimensionality increases, the data density decreases, as the average distance between objects increases with dimension~\cite{Jacox:2008:MSS:1366102.1366104}. Therefore, the average number of neighbors per point decreases significantly with dimensionality. While this demonstration is not representative of all possible scenarios, it shows two interesting computational problems. First, from Figure~\ref{fig:motivation}~(a), we see that the greatest response time (i.e., worst performance) occurs at 2 and 6 dimensions. In 2-D, there are many neighbors to consider per point, and while the R-tree index filters many of them, we must still perform costly Euclidean distance calculations between a large number of points to find those within the $\epsilon$ query distance.
Second, at 6-D there are very few neighbors within $\epsilon$ per point, unless $\epsilon$ is increased (Figure~\ref{fig:motivation}~(b)). However, the index search for 6-D data is more exhaustive in comparison to 2-D. Thus, index performance degenerates with increasing dimension, caused by the well-known curse of dimensionality~\cite{bellman1961,Durrant:2009:NNM:1558391.1558528,Volnyansky:2009:CDP:1637863.1638181,Kim2015}.

In this paper, we propose an algorithm that leverages GPU hardware to efficiently compute the distance similarity self-join.
This paper makes the following contributions:
\begin{itemize}[leftmargin=*]
\setlength{\itemsep}{1pt}
\setlength{\parskip}{0pt}
\setlength{\parsep}{0pt}

\item We present an efficient indexing strategy that bounds the search for nearby points and exploits the high memory bandwidth and parallelism afforded by the GPU. 
\item By carefully considering the domain decomposition provided by the index,  we provide a selection strategy that eliminates duplicate searches and distance calculations. We show that the performance of this optimization is dependent on GPU cache utilization. 
\item For self-joins that yield a large number of nearby neighbors which would exceed the memory capacity of the GPU, we exploit an efficient batching scheme between host and GPU to incrementally compute the entire self-join result.
\item We show that for a fixed dimension, our approach achieves similar performance gains on real-world and uniformly distributed synthetic datasets. 
\item Related to the above, we motivate that uniformly distributed data should represent a worst-case scenario for our grid-based indexing approach. Despite this, our approach is still significantly faster than the reference implementation using an R-tree across the 2--6 dimensions evaluated. Furthermore, our approach outperforms the parallel state-of-the-art implementation across nearly all experimental scenarios.
\end{itemize}
The paper is outlined as follows:
Section~\ref{sec:background} provides background information, Section~\ref{sec:probstatement} formalizes the problem, Section~\ref{sec:gpuindexing} presents our GPU indexing strategy, Section~\ref{sec:optimizedim} describes additional techniques used to improve self-join performance, Section~\ref{sec:expereval} outlines the methodology and experimental evaluation, and finally, we conclude the work in Section~\ref{sec:conclusions}.

\section{Background}\label{sec:background}
We consider several relevant categories of related work. The self-join problem is a special case of a join operation on two different sets of data points (or feature vectors). It is also similar to the problem of querying a database to find the subset of points whose values are within an $\epsilon$ distance from a query point. Therefore, works in these other areas are directly applicable to the self-join problem. All of these operations are typically supported by indexing data structures that are used to accelerate range queries on the input dataset.  Thus, since we use GPUs to improve performance in this work, we also consider related work that focuses on algorithmic transformations and GPU optimized indexes. We present an overview of each of these relevant areas below.

\noindent\textbf{Search-and-refine:}
Several works have studied the similarity-join problem on the CPU~\cite{Bohm:2000:HPC:354756.354832,Bohm2001,Arasu:2006:EES:1182635.1164206,Bayardo:2007:SUP:1242572.1242591,kalashnikov2013}. We focus on~\cite{Bohm:2000:HPC:354756.354832} as it uses the canonical search-and-refine strategy to compute the similarity self-join to accelerate clustering by calculating the neighbors of each point before clustering. 

The authors in~\cite{Bohm:2000:HPC:354756.354832} utilize the search-and-refine strategy and evaluate their approach using the R*-Tree~\cite{Beckmann:1990:RER:93597.98741} and X-tree~\cite{Berchtold:1996:XIS:645922.673502} to accelerate multidimensional searches on both 9-D and 64-D datasets.   They report that performing the self-join first, instead of a series of individual range queries in the instruction flow of the clustering algorithm can significantly improve clustering performance. Using the self-join over the iterative approach, they achieve significant performance gains in the context of out-of-core processing, which has different overheads than the in-memory processing examined here.  
This work shows that the self-join is used in other algorithms, and that indexes improve self-join performance.

\noindent\textbf{Employing Grids and Grid-based Indexes: }
Spatial indexes, such as the R*-Tree~\cite{Beckmann:1990:RER:93597.98741} mentioned above, are highly efficient at pruning the search space because the index is constructed based on the input data. Thus, regions with high data densities generate more data partitions (fewer partitions are needed in low density regions). This is similar to many other tree-based indexes such as k-d trees~\cite{bentley1975multidimensional} or quad trees~\cite{finkel1974quad}.

In contrast to these index-trees, grid-based methods make a static partitioning of the data, whereby some data partitions may contain over-densities of the data and other partitions may contain mostly empty space. Thus, the index is independent of the data distribution. One drawback of grids is that index-trees may be much faster at data retrieval on data distributions for which they are designed. Consequently, data-dependent index-trees have been the predominant index type for the search-and-refine strategy above. However, a benefit of grids is that they are fast to construct, and can be very space efficient, as their size is predominantly independent of the data distribution.

Grid-based approaches have been used to solve the similarity join problem on the CPU. The ``epsilon grid order''~\cite{bohm2001epsilon} overlays the physical space with a non-materialized $n$-dimensional grid, where each grid cell is of length $\epsilon$ (the query distance). To summarize their approach, finding points within $\epsilon$ is bounded to the points in the adjacent grid cells of the cell of a query point, and these cells can be pruned by determining how far away they lie based on the individual cell's $n$ dimensional coordinates.  Similarly, \cite{kalashnikov2013} advances an  epsilon grid order approach (Super-EGO) that uses the data distribution to inform pruning the search, and is  state-of-the-art for join operations. We compare our approaches to a multi-threaded implementation of Super-EGO in~\cite{kalashnikov2013}. 

\noindent\textbf{Indexes for the GPU: }
To reduce the number of distance calculations, several papers have advanced GPU-efficient indexes~\cite{Yang:2007:IGF:1363189.1363196,bohm2009,Zhang:2012:USH:2390226.2390229,Kim2015,Gowanlock2015,Gowanlock2016,KIM2018195}. A major question in this field is whether indexes suited for the CPU are efficient when implemented for the GPU. Since tree-based indexes require many branch instructions, a loss in parallel efficiency can occur on the GPU due to the SIMD architecture~\cite{Han:2011:RBD:1964179.1964184}. A key insight from~\cite{KIM2018195} is that it is preferable to have a less selective index that is very efficient on the GPU, rather than a more selective (tree-based) index that suffers from additional overheads and lower parallel efficiency. Consequently, we will motivate the use of a non-tree-based indexing scheme. A hierarchical grid index for the GPU that is designed for efficient searches in skewed data distributions is presented in~\cite{Yang:2007:IGF:1363189.1363196}. In our evaluation, we utilize both real-world and uniformly distributed data. However, future work includes examining skewed data in greater detail.

\noindent\textbf{GPU Similarity Joins: }
The similarity-join in~\cite{bohm2009} uses a directory structure to reduce the number of candidates that may be within $\epsilon$ of a query point. On 8-D datasets of up to 8 million points, the GPU outperforms the CPU algorithm when $\epsilon$ is selected such that each point has 1--2 average neighbors~\cite{bohm2009}. Given the input data and result set sizes, the self-join may not have exceeded the GPU's global memory capacity. We use batching to enable results that exceed GPU memory capacity. 

LSS~\cite{lieberman2008fast} advances a GPU similarity join for high-dimensional data that achieves impressive results. However, the target dimensionality is higher than that considered here.

\section{Problem Statement}\label{sec:probstatement}
The distance similarity self-join problem is described as follows.  Let $D$ be a database of points. Each point is denoted as $p_i$, where $i=1,\dots,|D|$. Each $p_i\in D$ has coordinates in $n$-dimensions, where each coordinate is denoted as $x_j$ where $j=1,\dots,n$, and $n$ is the number of dimensions of the point/feature vector. We use the following notation to denote a data point and associated coordinates: $p_i=(x_1,x_2,\dots,x_{n})$.

We find all points, $p_i \in D$, that are within the Euclidean distance, $\epsilon$, of each other (known as a range or distance query). As an example, given points $a \in D$ and $b \in D$, we say that the points are within $\epsilon$ when the distance function, $dist(a,b)\leq\epsilon$, where $dist(a,b)=\sqrt{\sum_{j=1}^n(a(x_j)-b(x_j))^2}$.  All processing occurs in-memory on the GPU and CPU.

\section{GPU Indexing}\label{sec:gpuindexing}
\subsection{Motivation: Utilizing a Grid}
There are a number of challenges that need to be overcome for efficient indexing on the GPU. Due to the constrained global memory capacity, it is crucial that the index structure is small. However, spatial index structures often encapsulate the entire space, and thus may retain information regarding empty space. While this may be feasible in low dimensionality (i.e., 2-D~\cite{Gowanlock2017}), it is intractable in higher dimensions.  In contrast to previous work in~\cite{Gowanlock2017}, we do not index empty cells.

Index-trees have been implemented for both the CPU~\cite{Guttman-R_tree} and GPU~\cite{Kim2015}.  However, the non-deterministic nature of tree-traversals leads to branch divergence, which is known to degrade GPU performance due to thread serialization~\cite{Han:2011:RBD:1964179.1964184}. Also, the access patterns of such tree-traversals may lead to poor cache efficiency.  In contrast, a search for nearby points in a grid can be bounded to adjacent cells (as discussed in the related work). Since the search is bounded, it is likely to have more regular memory access patterns in comparison to index-trees. As has been shown in other contexts, algorithms with regular memory access patterns on GPUs often yield significant performance gains over multi-core CPU algorithms~\cite{Burtscher2012}.

Because the grid is constructed independently of the data distribution (with the exception of encompassing the entire dataset in each dimension), the grid is well-suited for employing a work-avoidance strategy. As we demonstrate, employing a grid enables us to reduce the number of distance calculations and index searches by a factor of $\sim$2.

We utilize a grid-based index for the GPU because: $(i)$ the memory requirements of the index are small, which is important to maximize the space for other data; $(ii)$ bounded grid index searches have regular memory access patterns and thus less thread divergence, which would otherwise degrade performance; and, $(iii)$ the grid index presents an opportunity to reduce the total workload of the self-join.

\subsection{Index Properties}
 In this section, where appropriate, we use similar notation as previous work~\cite{Gowanlock2017} to describe our index. We denote $g_j$ as the properties of the index structure in the $j^{th}$ dimension.  We find the minimum and maximum values of each dimension across all points.  Using these values, we define as $g^{min}_j$ and $g^{max}_j$ as the minimum and maximum dimensions of our index grid, respectively. This defines the index range [$g^{min}_j$, $g^{max}_j$] where $g^{min}_j=\mathrm{min}_{p_i \in D}~x_j-\epsilon$ and $g^{max}_j=\mathrm{max}_{p_i \in D}~x_j+\epsilon$, for each dimension $j$. The range is appended by $\epsilon$ to avoid boundary conditions in grid cell lookup calculations.

\begin{figure*}[htp]
\centering
  \includegraphics[width=0.75\textwidth]{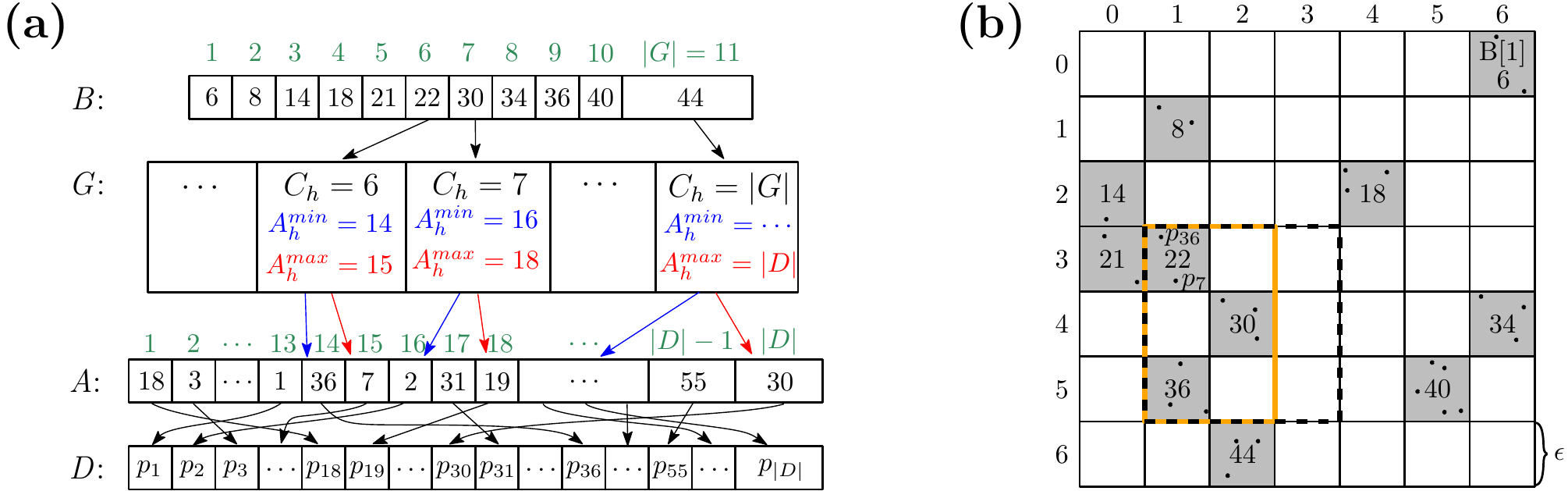}
    \caption{Example of our point indexing strategy.  (a) contains the indexing structure we employ and (b) illustrates the resulting grid of points in 2-D space.}
   \label{fig:GPU_index}
\end{figure*}

To generate an $n$-dimensional grid, we compute grid cells in each dimension, where the length of each cell is $\epsilon$. The number of cells in each dimension, $|g_j|$, is computed as $|g_j|=(g^{max}_j-g^{min}_j)/\epsilon$. For simplicity, we assume $\epsilon$ evenly divides the range of each dimension ($g^{max}_j-g^{min}_j$). This ensures that, for a query point in a cell, the search for neighboring points within $\epsilon$ will be constrained to the adjacent cells, bounding the search and regularizing the instruction flow.

If all cells were indexed, this would yield $\prod_{j=1}^{n} |g_j|$ grid cells. For even moderate dimensionality datasets, the total number of cells would make the space complexity intractable. Thus, in contrast to \cite{Gowanlock2017} we elect to only store non-empty grid cells in our grid index, denoted as $G$. The space occupied by the index is a function of the data density and distribution and not the total (hyper)volume defined by $\prod_{j=1}^{n}g^{max}_j-g^{min}_j$.

\subsection{Index Components}
In this section, we describe the components of the index shown in Figure~\ref{fig:GPU_index}~(a) and in the following section describe the index search using an example grid in Figure~\ref{fig:GPU_index}~(b). To minimize the space occupied by the index structure, we use several supporting components, enabling us to only store non-empty grid cells. When finding the adjacent grid cells of a point, we determine if a grid-cell exists based on the coordinates of a point and a grid cell lookup array, denoted as $B$. The number of non-empty grid cells is denoted as $|G|$, thus $|G|=|B|$; however, both sizes are dependent on the data distribution of $D$.  $B$ simply maps a cell's linearized coordinate computed from the cell's $n$-dimensional coordinates to its location in $G$.   To store the points in grid cells, we utilize a lookup array denoted as $A$. Each non-empty grid cell, denoted as $C_h$, $h\in\{1,\dots,|G|\}$, is stored as a linearized cell id, and contains the minimum and maximum ranges in $A$ that map to $D$. Thus, $C_h$ stores a range [$A_h^{min},A_h^{max}$]. Since $A$ provides a direct mapping between $D$ and the points inside each grid cell, $|A|=|D|$. A masking array is also computed to reduce the search space by filtering the cells as a function of the $n$-dimensional coordinates. We store only the coordinates of grid cells that are non-empty in each dimension. We denote this masking array as $M_j$, where $j\in\{1,\dots,n\}$.

\subsection{Index Search}
We illustrate searching the index. Figure~\ref{fig:GPU_index}~(b) shows a 2-D example grid, where non-empty cells are shaded and are labeled in lexicographic order. In the figure, there are 11 non-empty cells, thus $|B|=|G|=11$. Cell $C_h=7$ (linear id 30) contains a point $a\in D$, which requires calculating the neighbors within $\epsilon$ of the adjacent grid cells. We begin by storing the range in each dimension of the neighboring grid cells of $C_h=7$, in a range for each dimension, denoted as $O_j$, where $j=1,\dots,n$, which yields $O_1=[1,3]$ and $O_2=[3,5]$ in the respective dimensions (cells outlined with the black dashed lines in Figure~\ref{fig:GPU_index}~(b)).  We then compare $O_j$ to the masking array $M_j$. In the example, $M_1=[1,2]$ (in $j=1$, there are no points in the 3rd cell column), and $M_2=[3,5]$.  We then take $O_j\cap M_j$, which yields $[1,2]$, and $[3,5]$, respectively (cells outlined in orange in Figure~\ref{fig:GPU_index}~(b)). Then, with the ranges in each dimension, for each cell (6 in total), we compute the linear coordinate and binary search $B$ (Figure~\ref{fig:GPU_index}~(a)) to see if the linear coordinate exists. If so, we compute the distances to the points in the cell and filter the points with distances that are $>\epsilon$. For example, $C_6$ has linear coordinate 22, and the points in $D$ are found using $A$, as  $\{A[A_h^{min}],\dots,A[A_h^{max}]\}$ which yields $\{p_{36}, p_{7}\}$.

As an alternative to the lookup array $A$, one could store the points within each cell, $C_h$, thus mapping $G$ directly to $D$. However, this would require allocating a constant amount of memory per cell, the size of which would be the cell containing the greatest number of points. This would waste a significant amount of memory. Likewise, we could store all cells, including those that are empty, but that would increase $|G|$ exponentially with the number of dimensions, potentially causing the index to require more memory than the global memory capacity. With our strategy, the space complexity required by the index is $O(|B|+|G|+|A|)$, where $|B|=|G|$ and $|A|=|D|$.  Since our cells contain at least one point, our space complexity simplifies to $O(|D|)$.

The curse of dimensionality~\cite{bellman1961} is expected to reduce the pruning efficiency of index searches~\cite{Kim2015} in higher dimensions. The number of adjacent cells to a query cell is $3^n$, and this leads to a significant number of cells to check to see if they exist in $B$. However, since average density decreases in higher dimensions, fewer adjacent cells to a query cell will be non-empty, allowing us to ignore many neighbor cells.

\subsection{Global Memory GPU Kernel}
In this section we provide details of \selfjoinglobal, the main GPU kernel of our algorithm, with pseudocode provided in
Algorithm~\ref{alg:gpuselfjoinglobal}. We refer to Figure~\ref{fig:GPU_index} when illustrating the kernel, and
thus the kernel is presented in 2-D. For this kernel, we employ $|D|$ threads, where each thread considers a single point
and finds all neighbors within $\epsilon$ using the grid-based index. Threads do not utilize shared
memory in this kernel. The GPU kernel takes as input $D$, $A$, $G$, $B$, $M$, and, $\epsilon$.  Each thread begins
by getting its global thread id and checking to see if the global id is larger than $D$
(lines~\ref{algline1.globalID}--\ref{algline1.checkglobalID}).
The thread then stores its point in registers (line~\ref{algline1.getLocalPoint}) and computes index ranges of
adjacent cells in each dimension (line~\ref{algline1.adjcells}), corresponding to the black dashed
box in Figure~\ref{fig:GPU_index}~(b).  The ranges are then filtered in each dimension using $M$
(line~\ref{algline1.filteredcells}), resulting in the orange box in Figure~\ref{fig:GPU_index}~(b).  The thread
then iterates over these filtered ranges in each dimension (lines~\ref{algline1.loop1range}-\ref{algline1.loop2range})
to create $2$-D coordinates that are used to compute a linear coordinate for each non-empty neighbor cell
(line~\ref{algline1.linearid}).
In the example in Figure~\ref{fig:GPU_index}, the two loops generate the linearized ids: 22, 23, 29, 30,
36, 37 (orange outline). Each of these are searched in $B$, resulting in the non-empty cells 22, 30, and 36.
For each of these cells, the thread finds the points contained therein using array $G$ to get the
minimum and maximum ranges in $A$ (lines~\ref{algline1.lookupMin}--\ref{algline1.lookupMax}).
The thread then iterates over all points in this range (line~\ref{algline1.innerLoop}), computing the distance
from the query point (lines~\ref{algline1.calcResult}--\ref{algline1.ifResult}).
If the distance
computed is within $\epsilon$, then the result is stored as a key/value pair (line~\ref{algline1.addToResultSet}),
where the key is the query point id and the value is the point found to be within the $\epsilon$ distance. After the kernel's execution, we sort the key/value pairs, and transfer the result to the host.

Our example in Algorithm~\ref{alg:gpuselfjoinglobal} illustrates \selfjoinglobal in 2-D (i.e., $n=2$). When $n>2$, additional loops are required after the loops on lines~\ref{algline1.loop1range}--\ref{algline1.loop2range}, and the adjacent cell ids and filtered cell id ranges (lines~\ref{algline1.adjcells}--\ref{algline1.filteredcells}) also compute these ranges for the additional dimensions.

\begin{algorithm}
\caption{The \selfjoinglobal Kernel.}
\label{alg:gpuselfjoinglobal}
\begin{algorithmic}[1]

\begin{footnotesize}
\Procedure{\selfjoinglobal}{$D$, $A$, $G$, $B$, $M$, $\epsilon$}
\State gid $\leftarrow$ getGlobalId() \label{algline1.globalID}
\If {gid $\geq|D|$}{ \Return} \EndIf \label{algline1.checkglobalID}
\State resultSet $\leftarrow \emptyset$ \label{algline1.initialize}
\State point $\leftarrow$ $D$[gid] \label{algline1.getLocalPoint}
\State adjRangesArr[n] $\leftarrow$ getAdjCells(point)\label{algline1.adjcells}
\State filteredRngs[n] $\leftarrow$ maskCellRange($M$,adjRangesArr[n])\label{algline1.filteredcells}
\For {dim1 $\in$ filteredRngs[1].min, $\dots$, filteredRngs[1].max}\label{algline1.loop1range}
\For {dim2 $\in$ filteredRngs[2].min, $\dots$, filteredRngs[2].max}\label{algline1.loop2range}
\State linearID $\leftarrow$ getLinearCoord(dim1,dim2)\label{algline1.linearid}
\If {linearID $\in$ $B$}\label{algline1.lookupB}
\State lookupMin $\leftarrow$ $A[G$[linearID].min$]$ \label{algline1.lookupMin}
\State lookupMax $\leftarrow$ $A[G$[linearID].max$]$ \label{algline1.lookupMax}
\For {candidateID~$\in$~\{lookupMin,$\dots$,lookupMax\}} \label{algline1.innerLoop}
\State result $\leftarrow$ calcDistance(point, $D$[candidateID]) \label{algline1.calcResult}
\If {result $\leq \epsilon$} \label{algline1.ifResult}
\State \textbf{atomic:} resultSet $\leftarrow$ resultSet $\cup$ result \label{algline1.addToResultSet}
\EndIf
\EndFor

\EndIf

\EndFor
\EndFor

\EndProcedure
\Return
\end{footnotesize}
\end{algorithmic}
\end{algorithm}

\section{Optimizations}\label{sec:optimizedim}
The indexing scheme presented in Section~\ref{sec:gpuindexing} provides a general method
for us to efficiently solve the self-join problem for datasets ranging from 2 to 6 dimensions.
In this section, we discuss optimizations that further improve the performance of our self-join
algorithm on the GPU.  Since this work focuses on datasets from 2--6 dimensions, we focus on
optimizations that address the challenges of processing low dimensional data.

\subsection{Batching the Result Set}
\label{sec:batching}
Batching schemes are required by many GPU-accelerated approaches that process large volumes of data. In low dimensionality, the hyper-volume of the space is small in comparison to high dimensionality, so more points co-occur in the same region, leading to larger result set sizes (e.g., Figure~\ref{fig:motivation}). Since there are potentially a large number of points within the $\epsilon$ distance, a bottleneck is the data transfer of the result set between the GPU and host. We leverage the work of~\cite{Gowanlock2017} that calculated the $\epsilon$-neighborhood of 2-D points on the GPU for use in the DBSCAN clustering algorithm~\cite{ester1996density}. In our context, without incrementally batching the result set it may not be possible to process low-dimensional data in a single kernel invocation as the result set may exceed global memory capacity. Furthermore, the batching scheme allows us to overlap GPU computation with bidirectional data transfers between host and GPU. Thus, even for workloads that would not overflow the GPU's global memory capacity, it is more efficient to use the batching scheme. In all experiments, the minimum number of batches is set to 3.

\subsection{Avoiding Duplicate Calculations }
Euclidean distance is reflexive, so if $p\in D$ is within $\epsilon$-distance of $q \in D$, then $q\in D$ is within $\epsilon$ of $p\in D$.
Thus, if we can evaluate pairs of neighboring points only \emph{once}, we can report both corresponding ordered pairs
and eliminate redundant work.  We present a general method of accomplishing this for $n$ dimensions, which we call
``uni-directional comparison'' (\unicomp).

\begin{figure}[ht]
  \centering
  \includegraphics[width=0.35\textwidth]{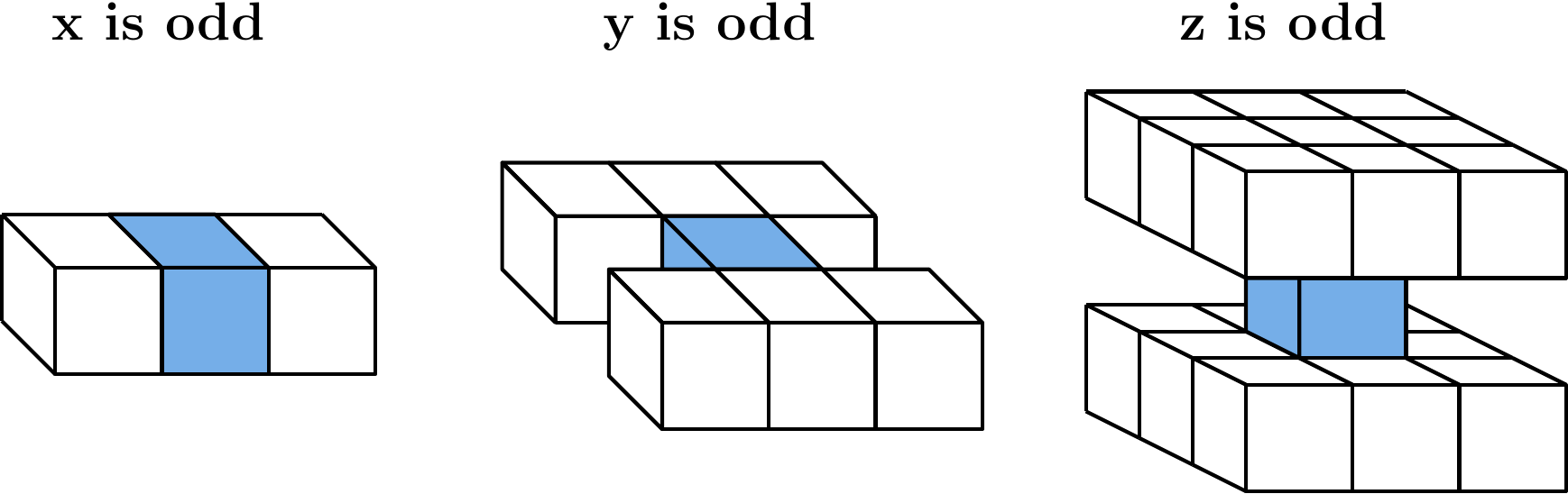}
  \caption{Illustration of the \unicomp optimization.
The blue cell is our source cell and the white cells drawn around it are those we want to evaluate
if the source cell index at the particular dimension is odd.}
  \label{fig:uni-dims}
\end{figure}

We consider each dimension separately and, for each point contained in a cell that has an \emph{odd} coordinate for the given
dimension, we evaluate points in a specific set of neighbor cells. Figure~\ref{fig:uni-dims} illustrates which neighbor
cells we want to evaluate when the query point is located in a (blue) cell with an odd index in each of the first 3 dimensions.

In the example of $n=3$ ($x$, $y$, and $z$ coordinates), we first consider the $x$-index. If cell $C_a$ containing the source point has an \emph{odd} $x$-index,
we evaluate the neighbor cells that differ by $x$-index but share the same
$y$ and $z$ indices as $C_a$.  If the cell has an even $x$-index,  we do nothing.  We then consider the $y$-index: if odd, we evaluate
all neighbor cells that differ by $y$ but share the $z$-index as $C_a$.  Finally, we consider $z$; if odd, we evaluate all neighbor cells
with $z$-index that differs from $C_a$.
Pseudocode illustrating this process for 3 dimensions ($x$, $y$, and $z$) is presented in 
Algorithm~\ref{alg:unicomp}.

\begin{algorithm}
\caption{The \unicomp access pattern 3 dimensions}
\label{alg:unicomp}
\begin{algorithmic}[1]

\begin{footnotesize}
\Procedure{\unicomp3D}{$point$, $C_a$, $filteredRngs$, $B$}

\If {$C_a$.x is odd}
\For {x $\in$ filteredRngs[1]}
\If {x $\neq$ $C_a$.x}
\State linearID $\leftarrow$ getLinearCoord(x, $C_a$.y, $C_a$.z)
\If {linearID $\in$ $B$}
\State ComparePoints(point, linearID)
\EndIf
\EndIf
\EndFor
\EndIf
\If {$C_a$.y is odd}
\For {x $\in$ filteredRngs[1]}
\For {y $\in$ filteredRngs[2]}
\If{y $\neq$ $C_a$.y}
\State linearID $\leftarrow$ getLinearCoord(x, y, $C_a$.z)
\If {linearID $\in$ $B$}
\State ComparePoints(point, linearID)
\EndIf
\EndIf
\EndFor
\EndFor
\EndIf
\If {$C_a$.y is odd}
\For {x $\in$ filteredRngs[1]}
\For {y $\in$ filteredRngs[2]}
\For {z $\in$ filteredRngs[3]}
\If{z $\neq$ $C_a$.z}
\State linearID $\leftarrow$ getLinearCoord(x, y, z)
\If {linearID $\in$ $B$}
\State ComparePoints(point, linearID)
\EndIf
\EndIf
\EndFor
\EndFor
\EndFor
\EndIf
\EndProcedure
\Return
\end{footnotesize}
\end{algorithmic}
\end{algorithm}

We apply this pattern when considering which neighbor cells to evaluate for each point $p$.   Whenever a point $q$ is found within the $\epsilon$
distance of $p$, we add both ($p$, $q$) and ($q$, $p$) to our result set.  \unicomp reduces both the index search overhead
(cell evaluations) and Euclidean distance calculations roughly by a factor of two.  Thus, we expect the \unicomp optimization to improve overall performance by this factor.

\section{Experimental Evaluation}\label{sec:expereval}

\subsection{Datasets}\label{sec:datasets}
We utilize both real and synthetic datasets to evaluate the performance of our approaches.
Our synthetic datasets \datasetsynParen assume data is uniformly distributed and independent in each dimension. As will be elaborated on in the experimental evaluation, uniformly distributed data represents a worst-case scenario for our GPU grid indexing scheme because it maximizes the number of non-empty cells, leading to higher search overhead in comparison to datasets with regions of higher density, relative to the average data density.  
We generate datasets in multiple dimensions with 2 and 10 million points to observe how performance varies as
a function of data density and dimensionality. Each data point is represented as a 64-bit floating point
value uniformly distributed in the range [0,100] in each dimension. 
Additionally, we evaluate performance using the following real-world datasets. The \datasetgeo dataset contains the latitude/longitude and the total electron content in the ionosphere~\cite{datasetsTEC}. We use this dataset in either 2 or 3 dimensions: in 2-D, we use the coordinates of the point, and in 3-D we include the total electron content value. The \datasetsdss datasets are galaxies from the Sloan Digital Sky Survey, data release
12~\cite{2015ApJS..219...12A}. Galaxies are represented in 2-D, spanning  redshift ($z$) of $0.30 \leq z\leq 0.35$. A summary of dataset properties are outlined in Table~\ref{tab:datasets}.

\begin{table}[htp]
\centering
\begin{footnotesize}
\caption{Dataset, data points, $|D|$, and dimension, $n$.}\label{tab:datasets}
\begin{tabular}{|c|c|c|c|c|c|} \hline
Dataset &$|D|$ & $n$ & Dataset &$|D|$ & $n$\\ \hline
\multicolumn{6}{|c|}{\datasetsyn}\\ \hline
\dsynthaa &$2\times10^6$&2&\dsynthba&$10\times10^6$&2\\ \hline
\dsynthab &$2\times10^6$&3&\dsynthbb&$10\times10^6$&3\\ \hline
\dsynthac &$2\times10^6$&4&\dsynthbc&$10\times10^6$&4\\ \hline
\dsynthad &$2\times10^6$&5&\dsynthbd&$10\times10^6$&5\\ \hline
\dsynthae &$2\times10^6$&6&\dsynthbe&$10\times10^6$&6\\ \hline
\multicolumn{6}{|c|}{Real World: \datasetgeo, \datasetsdss}\\\cline{1-6}
\dgeoaa & 1,864,620& 2 & \dgeoad & 5,159,737& 2\\ \hline
\dgeoba & 1,864,620& 3 & \dgeobd & 5,159,737& 3\\ \hline 
\sdssa&$2\times10^6$&2 & \sdssc&15,228,633&2\\ \hline
\end{tabular}
\end{footnotesize}
\end{table}

\subsection{Experimental Methodology}
The GPU code is written in CUDA~\cite{CUDA} and executed on an nVIDIA TITAN X (Pascal architecture) GPU with 12 GiB of global memory. All C/C++ host code is compiled with the GNU compiler with the O3 optimization flag.  The platform has $2\times 16=32$ Intel E5-2683 v4 2.1 GHz CPU cores. The self-join CUDA kernel (Algorithm~\ref{alg:gpuselfjoinglobal}) is configured to run with 256 threads per block, and uses 64-bit double precision floats. All results are averaged over 3 trials. We denote our GPU approach as \gpu.

\noindent \textbf{Reference Implementation (\rtree): } We compare the performance of \gpu to a sequential reference implementation using an R-tree~\cite{Guttman-R_tree} index, denoted as \rtree. The performance of the R-tree search is sensitive to the insertion order of the data, and co-located data should be inserted together (e.g., using a Hilbert curve~\cite{Kamel:1993:PR:170088.170403}). Thus, in all experiments, we first sort the data into bins of unit length in each dimension. This ensures that internal nodes of the R-tree do not encompass too much empty space. Since we focus on index search performance, we omit index construction time. We note, however, that inserting points into the grid in \gpu requires far less work than constructing the R-tree.

As discussed in Section~\ref{sec:intro}, the search-and-refine strategy that is reflected in \rtree is efficient in low dimensionality; however, index search performance degrades in higher dimensions. For this reason, \emph{we only focus on dimensions 2--6}. 

\noindent \textbf{State-of-the-art (\ego): } The Super-EGO algorithm~\cite{kalashnikov2013} for the CPU performs fast self-joins on multidimensional data. It has been shown to outperform many other join algorithms on both low and high dimensional data and is considered state-of-the-art. As such, we compare our approach to the multi-threaded version of Super-EGO~\cite{kalashnikov2013}, using 32 threads on our 32 core platform. We execute the algorithm using 32-bit floats (execution with 64-bit floats failed). However, our GPU implementation uses 64-bit floats; therefore, the \ego algorithm is expected to be slower if it were executed with 64-bit floats (e.g., more cache misses would occur). Furthermore, \ego normalizes all data in the range [0,1] in each dimension. We modified our datasets accordingly, but in our figures, we show the non-normalized value of $\epsilon$ so that we can compare results. We validated consistency between our implementations by comparing the total number of neighbors within $\epsilon$. In all time measurements, we use the total time to ego-sort and join. We are grateful to D. Kalashnikov for making his code publicly available.

\noindent \textbf{GPU Brute Force Implementation: } As dimensionality increases, there is increased index search overhead in any self-join operations that use an index. Thus, at some dimension, a brute force nested loop join $O(|D|^2)$ algorithm~\cite{kalashnikov2013} that compares all points to each other is expected to be more efficient than using an index. 
Since this brute force approach compares all pairs of points, it is independent of $\epsilon$.
We also compare to a parallel brute force implementation on the GPU. The kernel simply uses $|D|$ threads, where each thread is assigned one point to compare to each of the other points in the dataset. We exclude the time to transfer the result set back to the host, and thus only execute a single kernel invocation. This represents a lower bound on the brute force implementation. Since the performance of the brute force algorithm is not dependent on $\epsilon$, we only run the brute force algorithm for a single value of $\epsilon$ on a given dataset. This GPU brute force implementation demonstrates that the performance gains of our approaches are \emph{not} due to the high throughput of the GPU.

\subsection{Results}
As the search distance, $\epsilon$, increases, the performance of the self-join degrades for two reasons: $(i)$ more comparisons are required to determine if two points are within $\epsilon$ of each other; and, $(ii)$ the performance of index searches degrade as the search volume increases. Our grid-based index is less susceptible to the effect in $(ii)$ above because the search is limited to adjacent grid cells of a given query point.  

\begin{figure}[!t]
\centering
\subfigure[\dgeoaa]{
        \includegraphics[width=0.22\textwidth, trim={0.6cm 0 0.6cm 0}]{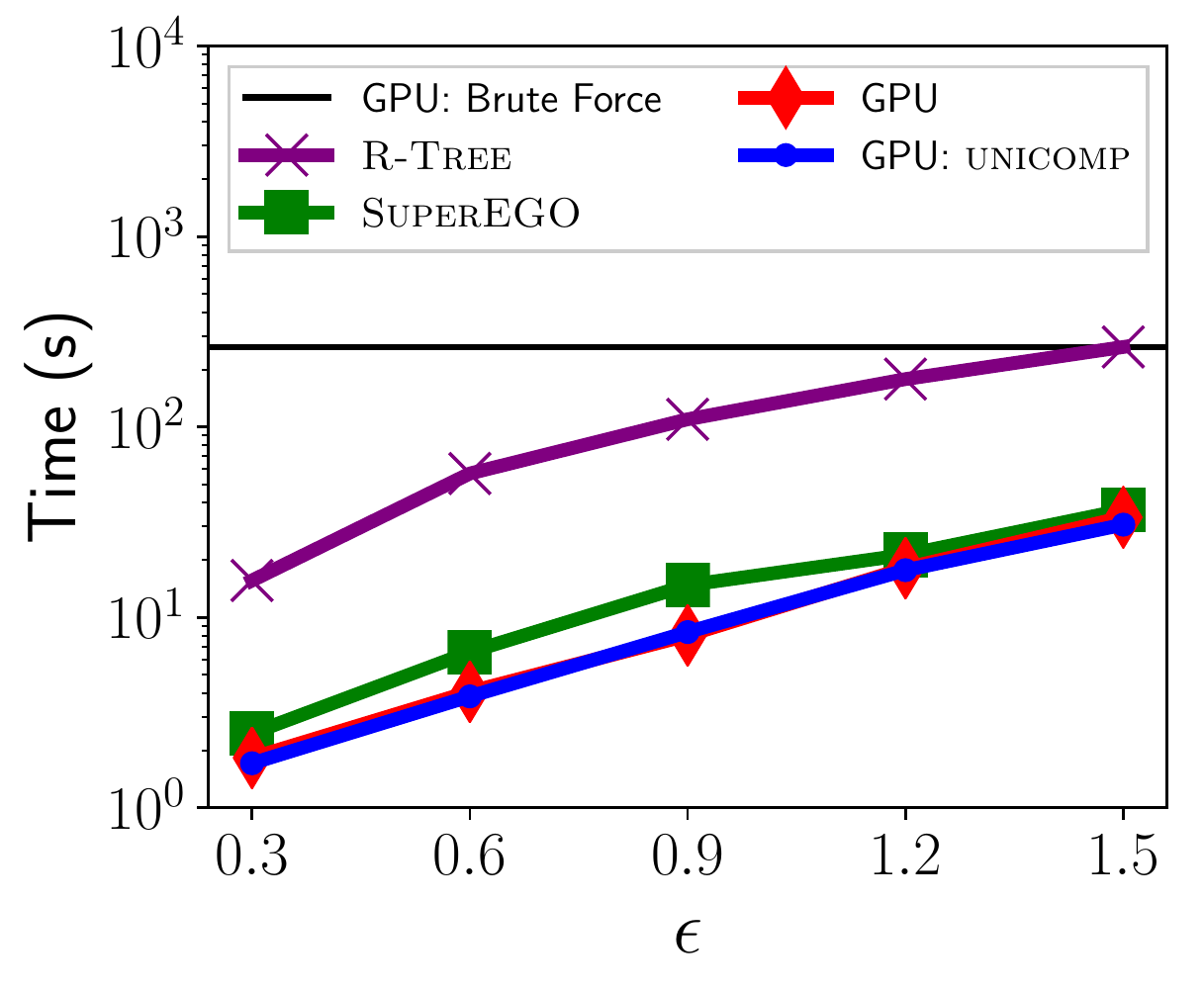}
    }
\subfigure[\dgeoad]{
        \includegraphics[width=0.22\textwidth, trim={0.6cm 0 0.6cm 0}]{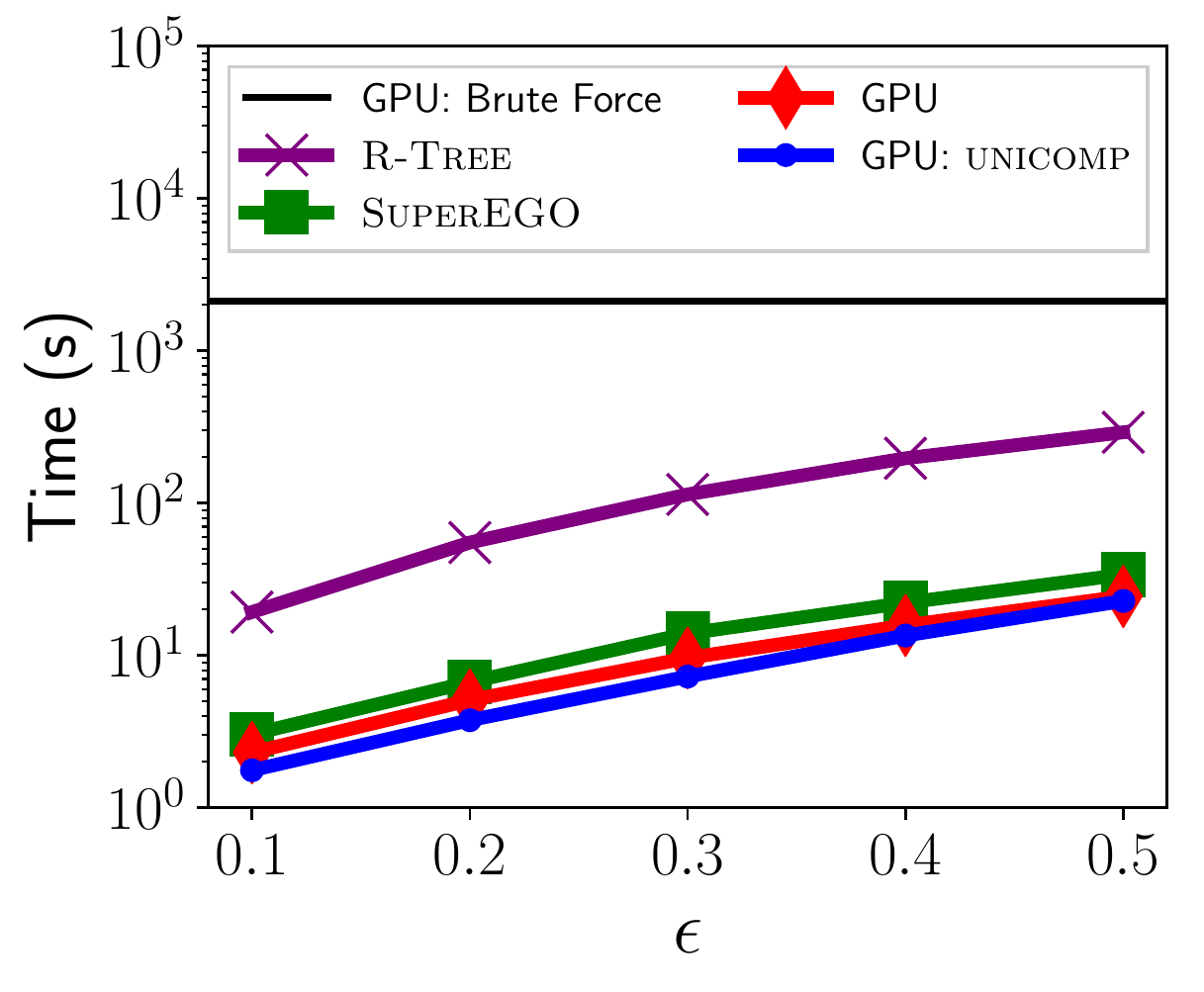}
    }

\subfigure[\sdssa]{
        \includegraphics[width=0.228\textwidth, trim={0.6cm 0 0.6cm 0}]{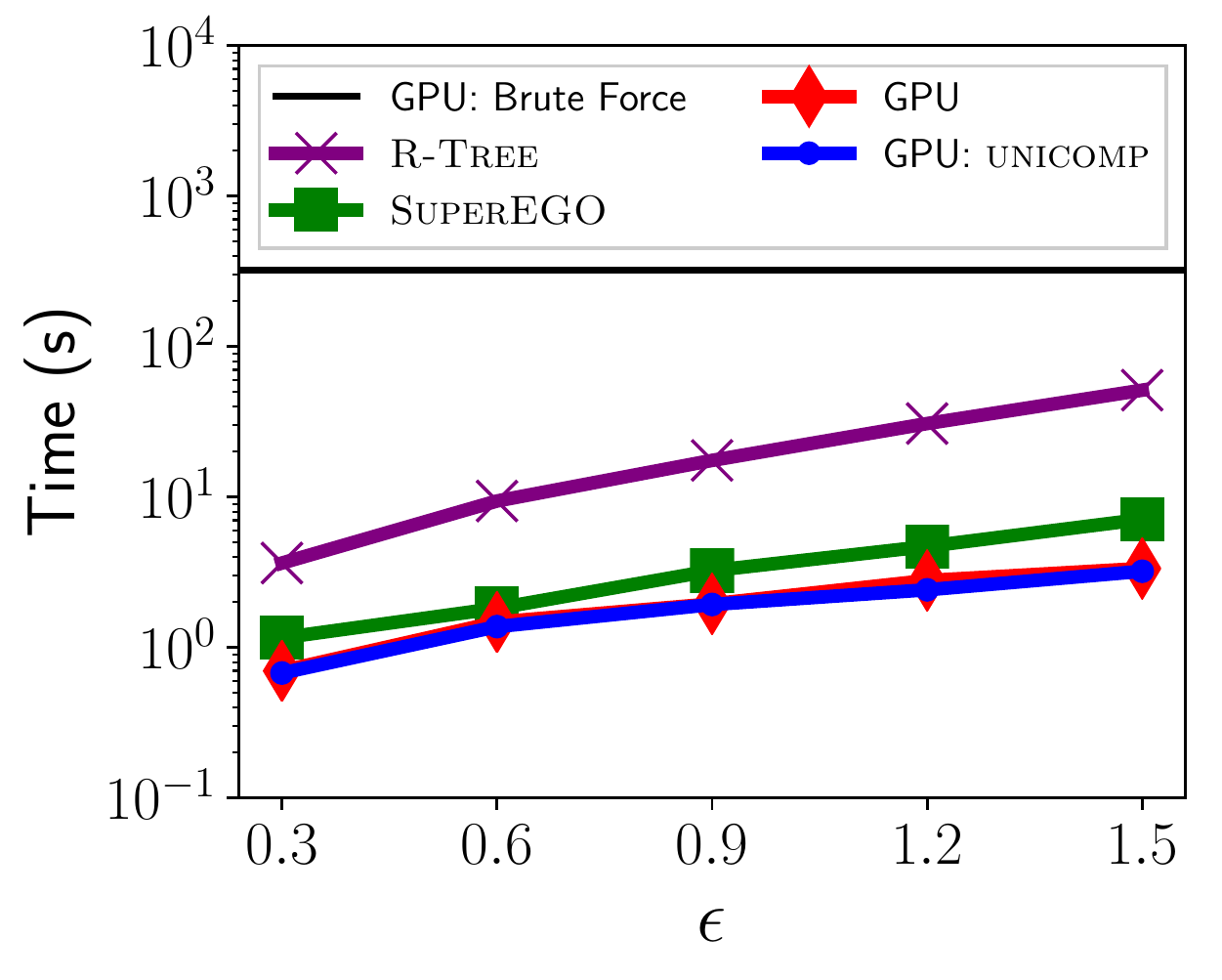}
    }
\subfigure[\sdssc]{
        \includegraphics[width=0.22\textwidth, trim={0.6cm 0 0.6cm 0}]{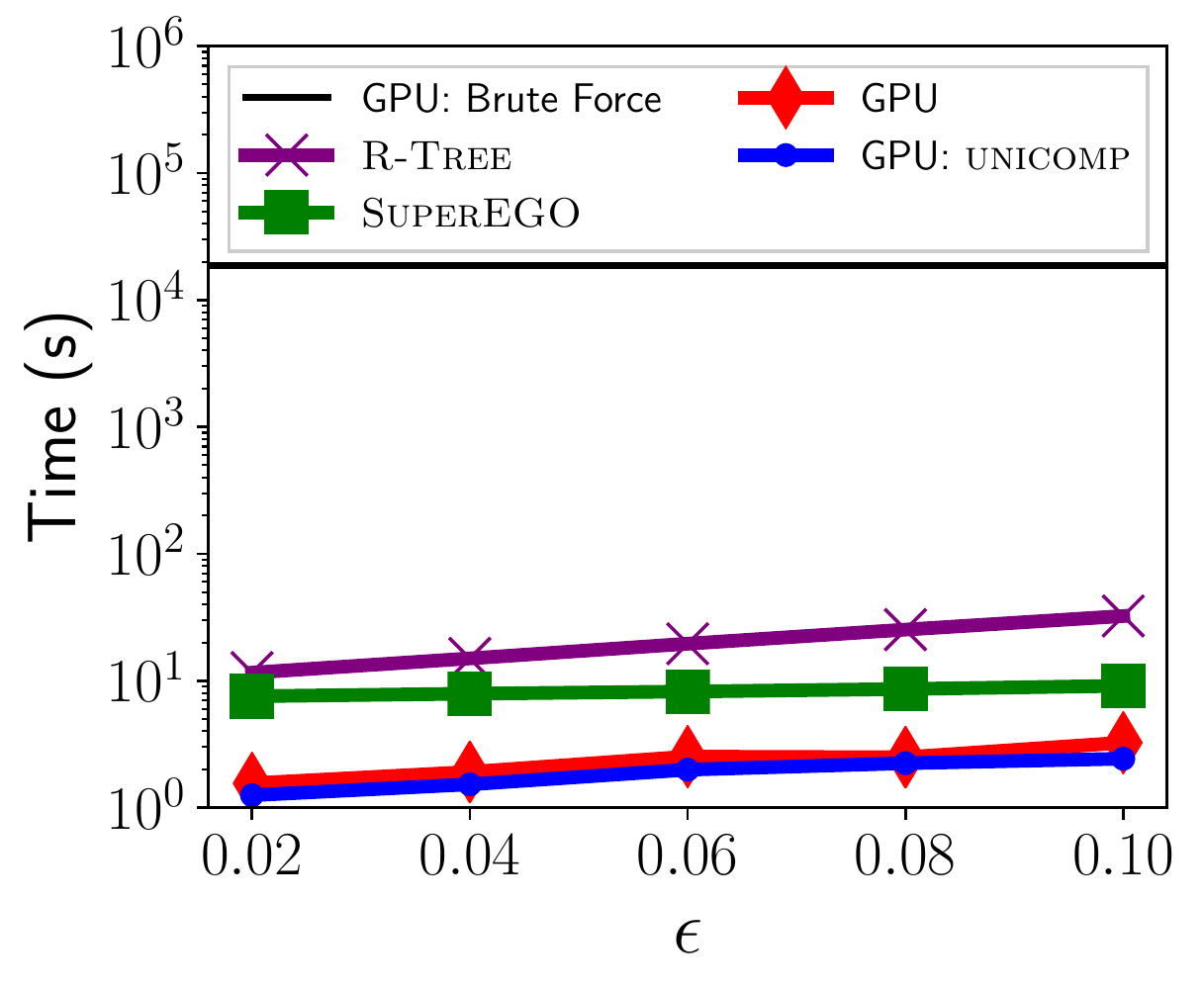}
    }

\subfigure[\dgeoba]{
        \includegraphics[width=0.22\textwidth, trim={0.6cm 0 0.6cm 0}]{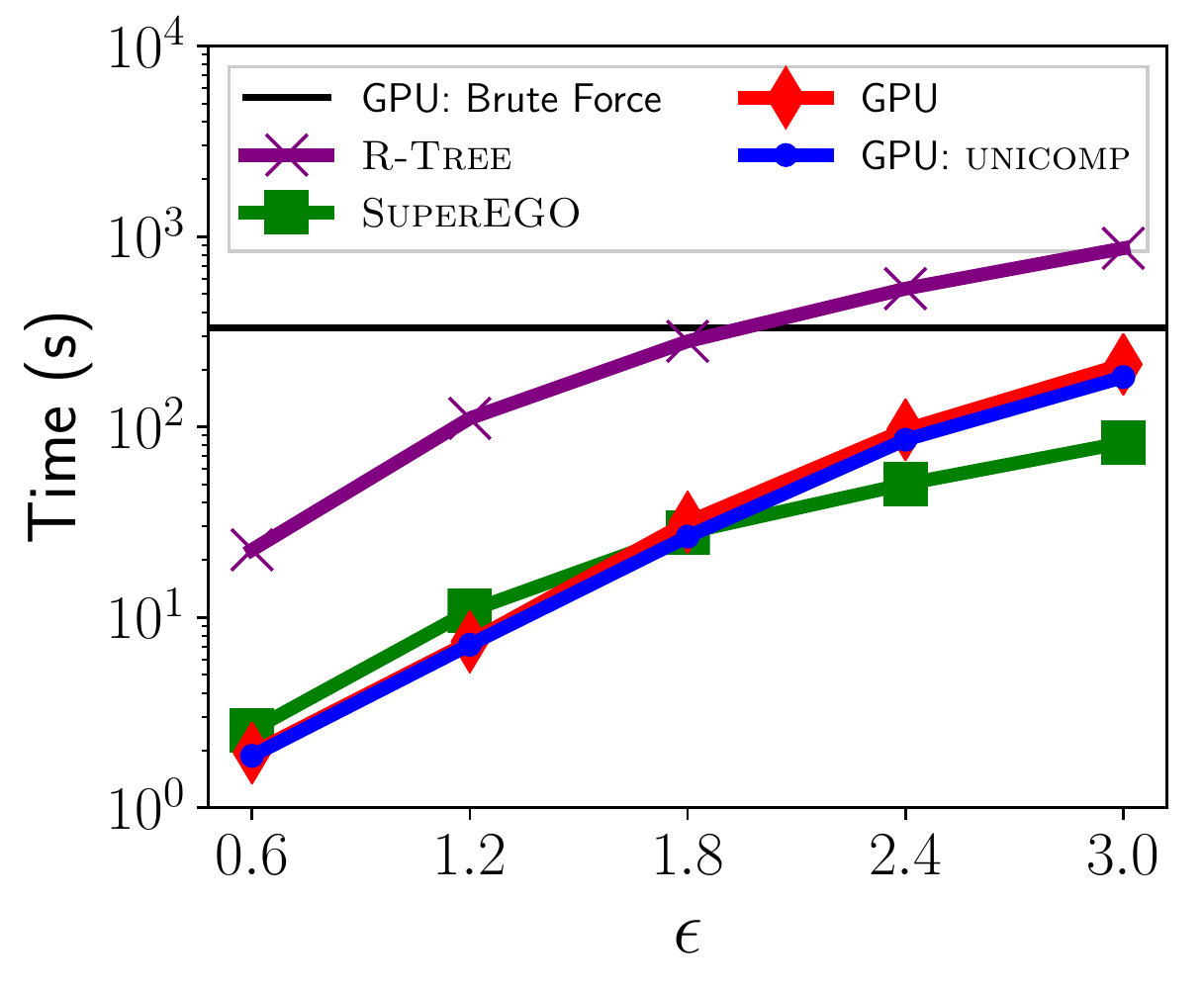}
    }
\subfigure[\dgeobd]{
        \includegraphics[width=0.22\textwidth, trim={0.6cm 0 0.6cm 0}]{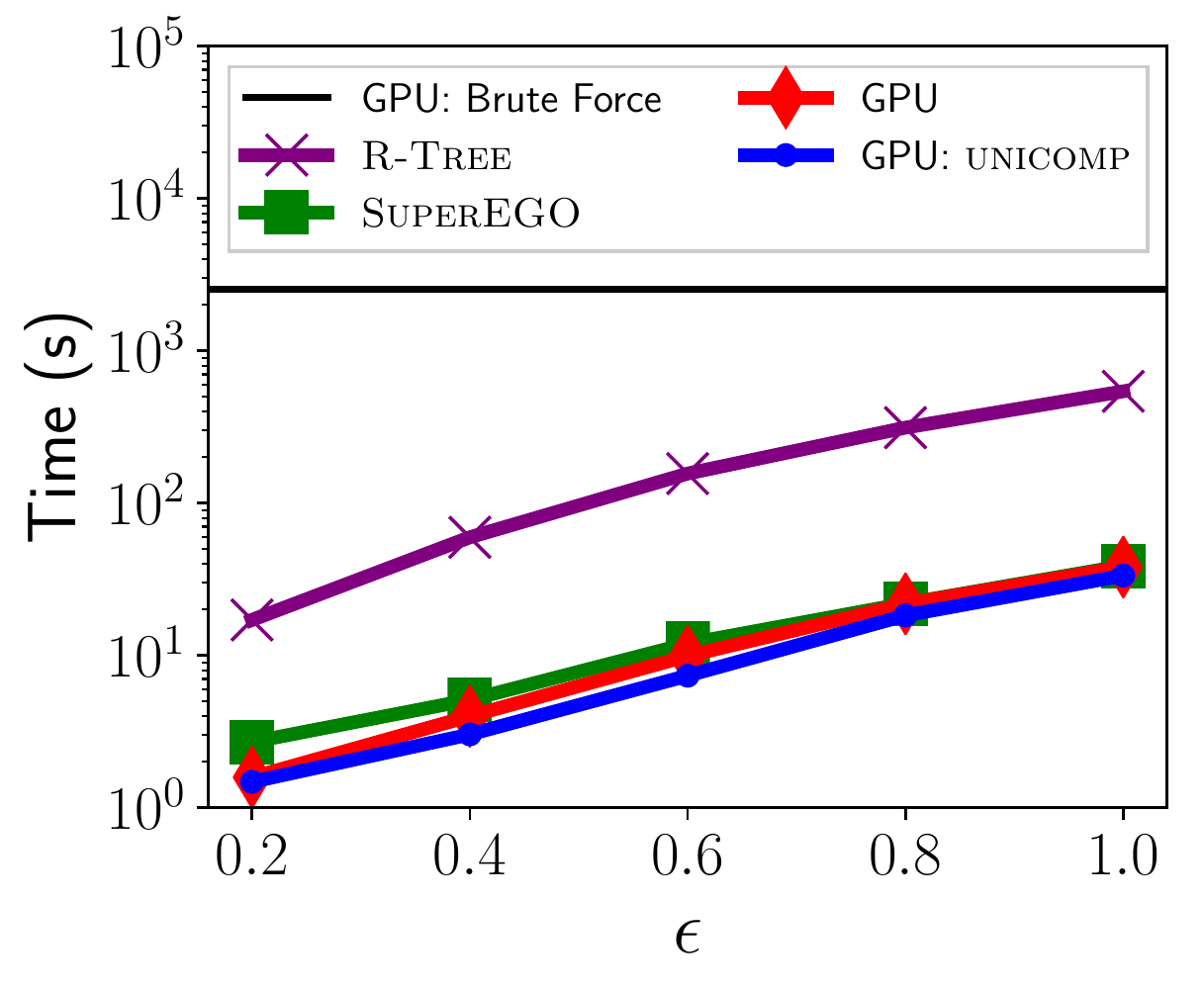}
    }    
    \caption{Response time vs. $\epsilon$ on the real-world datasets, \datasetgeo (a, b, e, f) and \datasetsdss (c, d). The \datasetsdss datasets are in 2-D and \datasetgeo span 2--3 dimensions.}
   \label{fig:real_world}
\end{figure}

\noindent \textbf{Real-world datasets: } Figure~\ref{fig:real_world} plots the response time vs. $\epsilon$ for the real-world datasets, which span 2--3 dimensions. The figure is plotted on log scale so that we can observe differences in response times across multiple orders of magnitude (we summarize speedup in a subsequent figure). Across all datasets, \gpu outperforms \rtree. Furthermore, \gpu outperforms the state-of-the-art algorithm, \ego (executed in parallel with 32 threads) across most scenarios. We find that on \dgeoba when $\epsilon>1.8$, \ego outperforms \gpu with \unicomp. However, on the largest dataset, \sdssc (containing 15 million points), \gpu with \unicomp achieves up to $\sim$6$\times$ speedup over \ego.
Although \gpu is faster than the brute force method for all of our experiments, on
the \dgeoba dataset (Figure~\ref{fig:real_world}~(e)), brute force outperforms \rtree
at $\epsilon>1.8$.  
This is because the $\epsilon$ values are quite large in this experiment, increasing the number of distance calculations and diminishing the efficacy of the index.

\begin{figure*}[!thp]
\centering
\subfigure[\dsynthaa]{
        \includegraphics[width=0.181\textwidth, trim={0.7cm 0 0.7cm 0}]{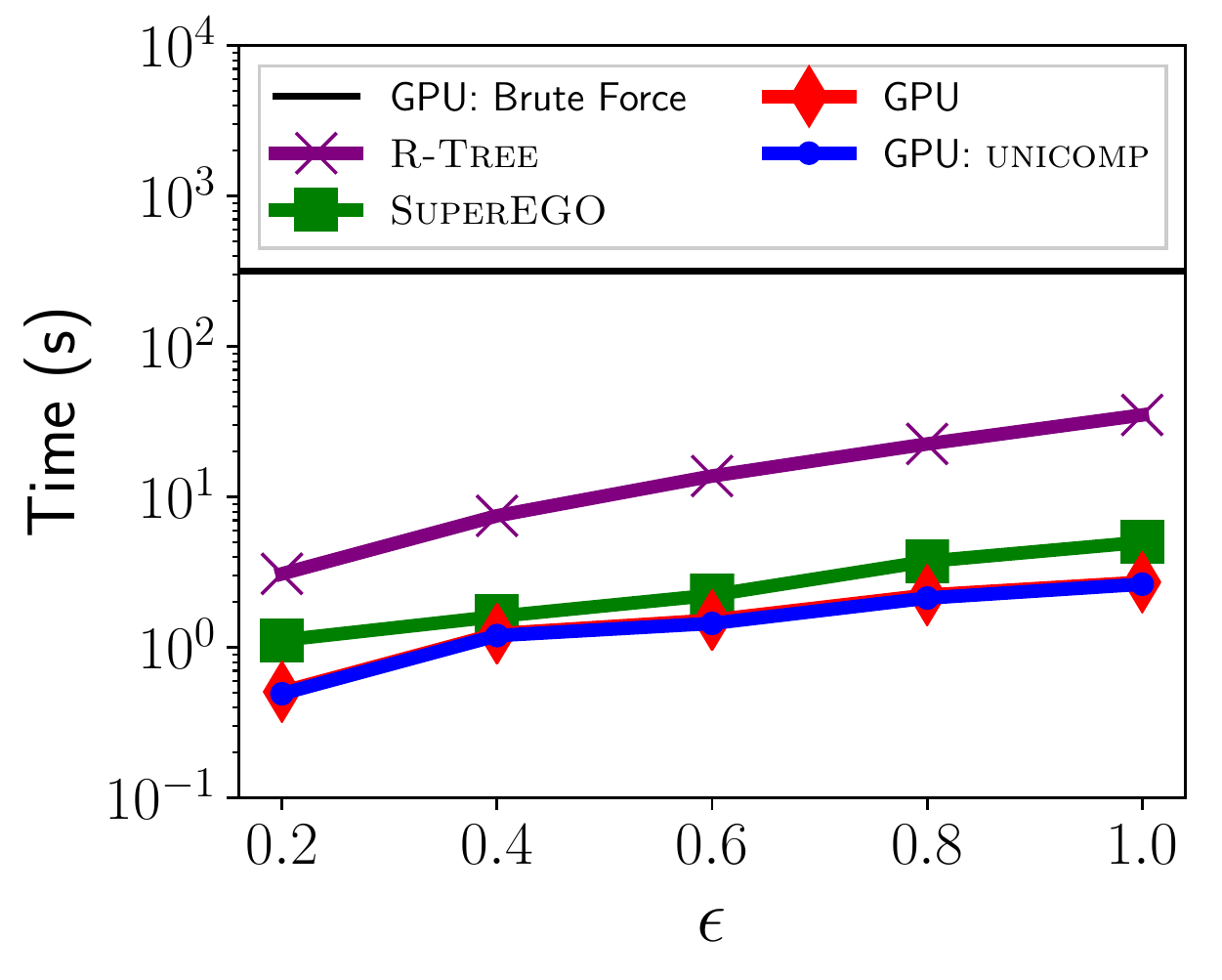}
    }
\subfigure[\dsynthab]{
        \includegraphics[width=0.181\textwidth, trim={0.7cm 0 0.7cm 0}]{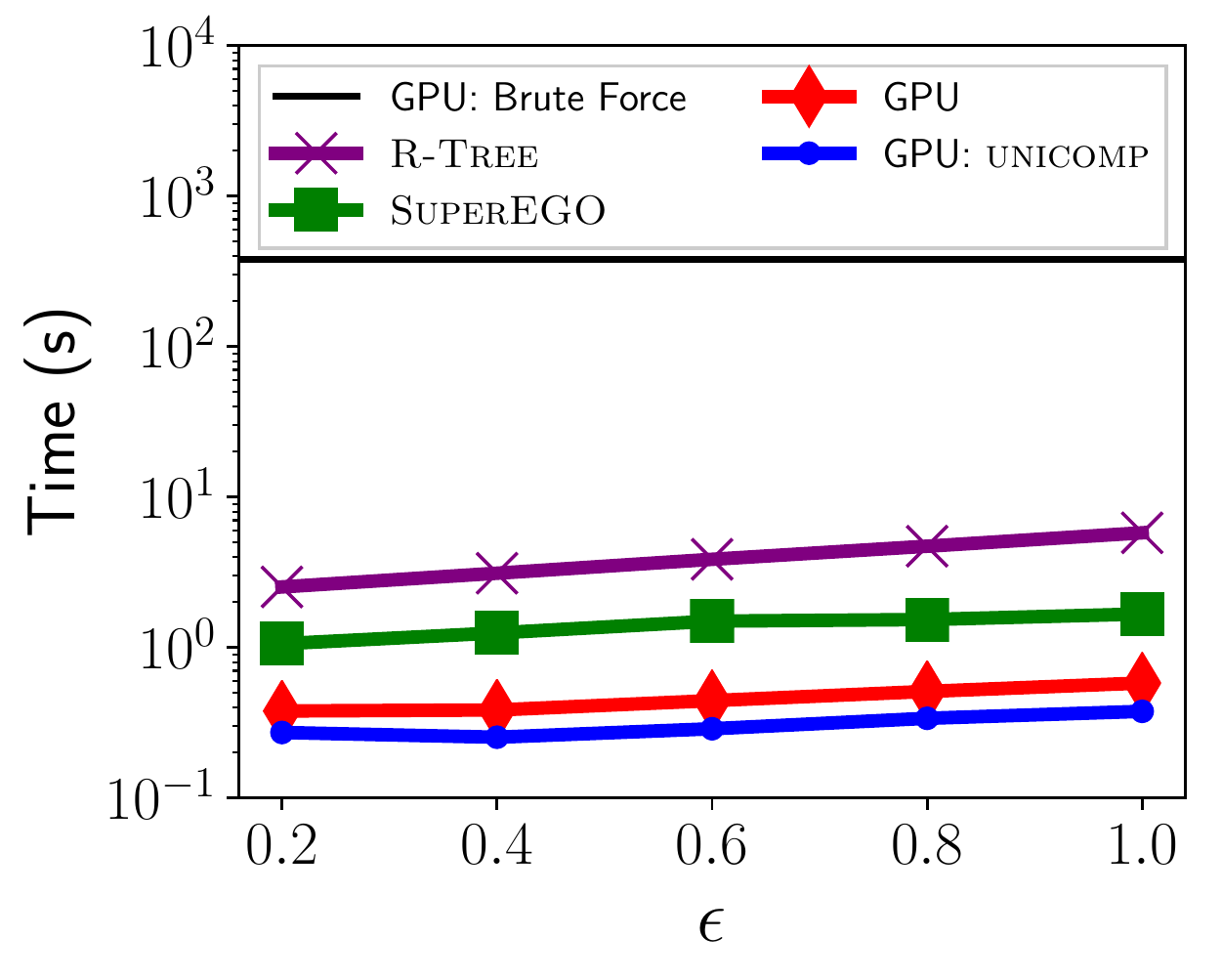}
    }
\subfigure[\dsynthac]{
        \includegraphics[width=0.181\textwidth, trim={0.7cm 0 0.7cm 0}]{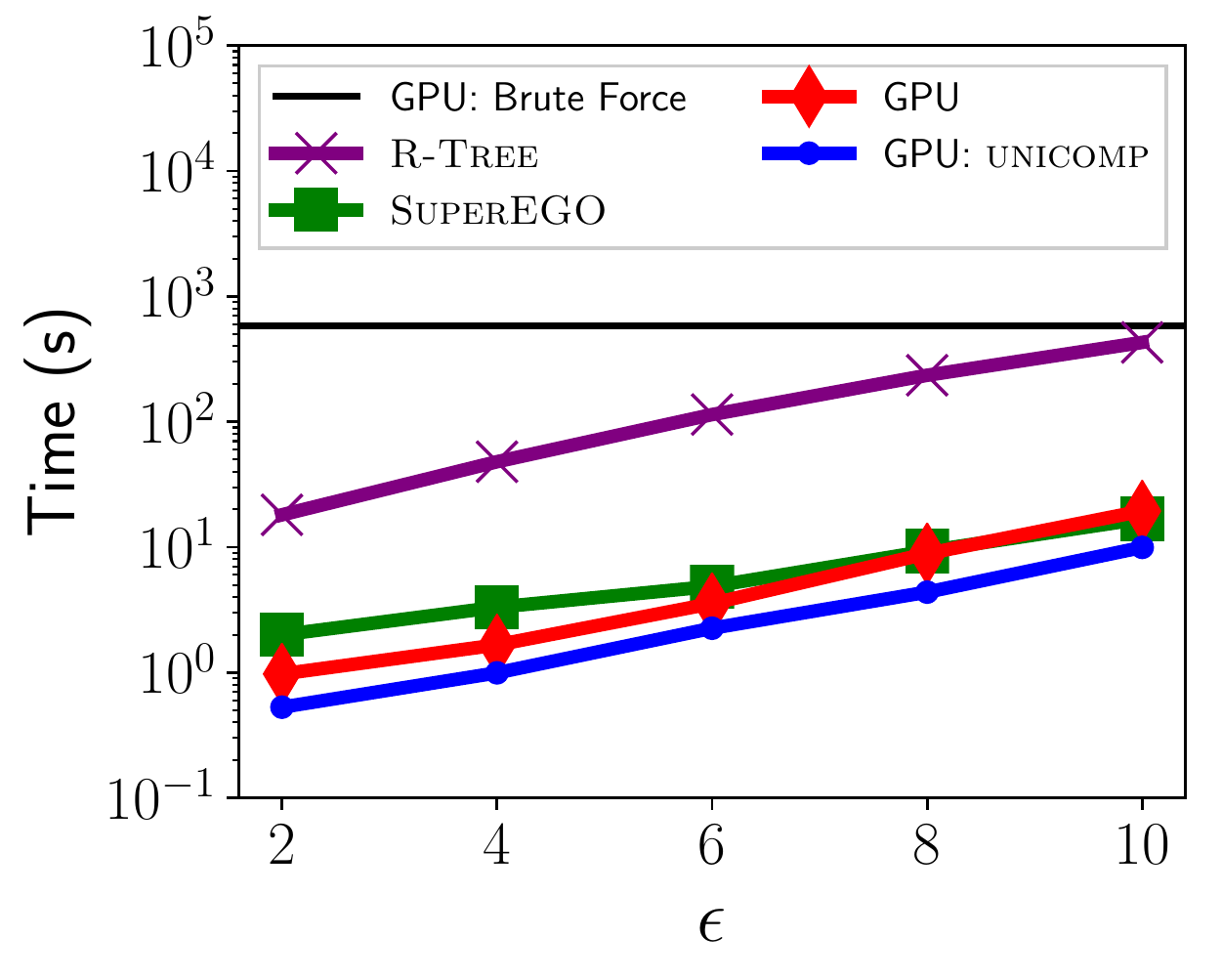}
    }
\subfigure[\dsynthad]{
        \includegraphics[width=0.181\textwidth, trim={0.7cm 0 0.7cm 0}]{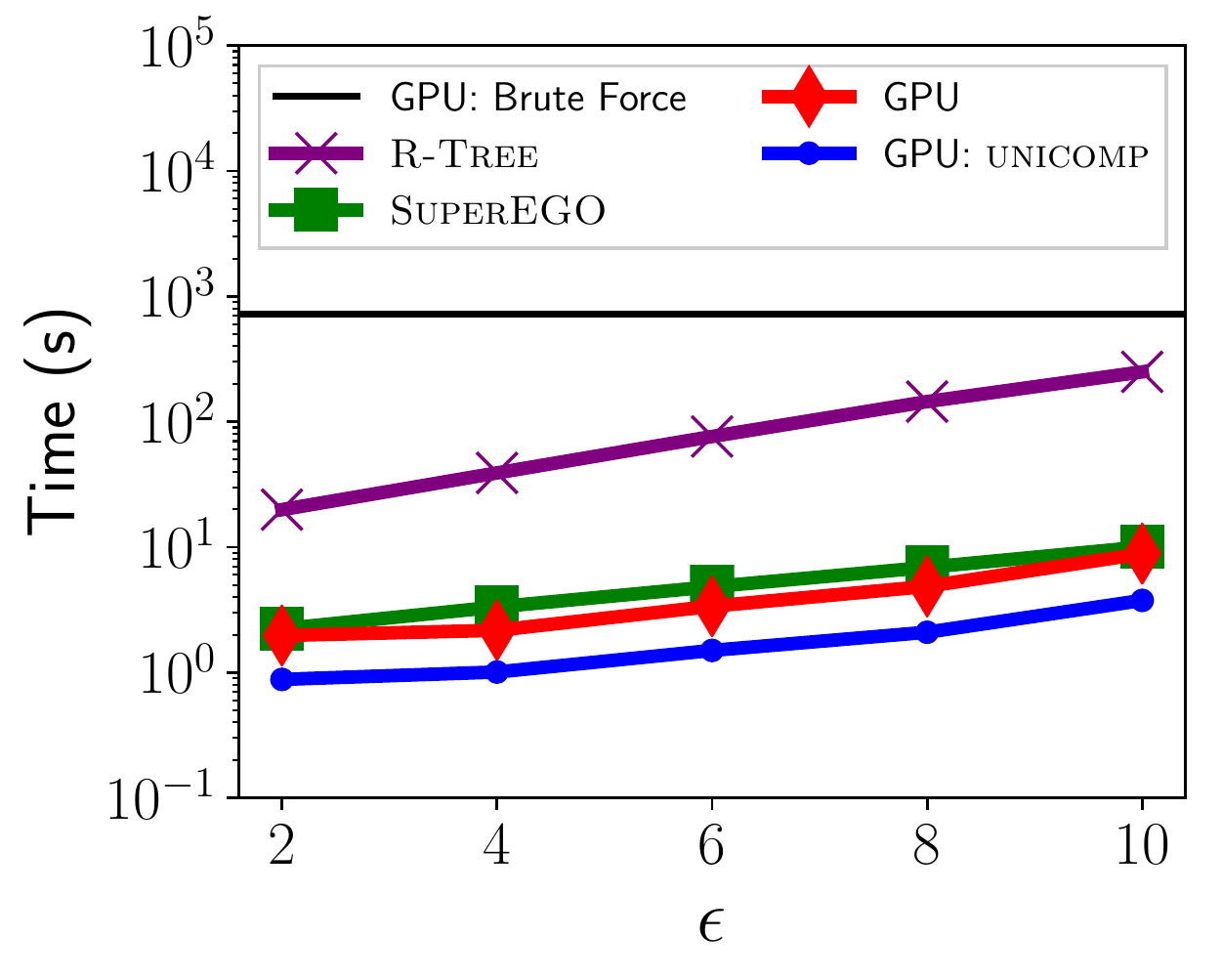}
    }
\subfigure[\dsynthae]{
        \includegraphics[width=0.181\textwidth, trim={0.7cm 0 0.7cm 0}]{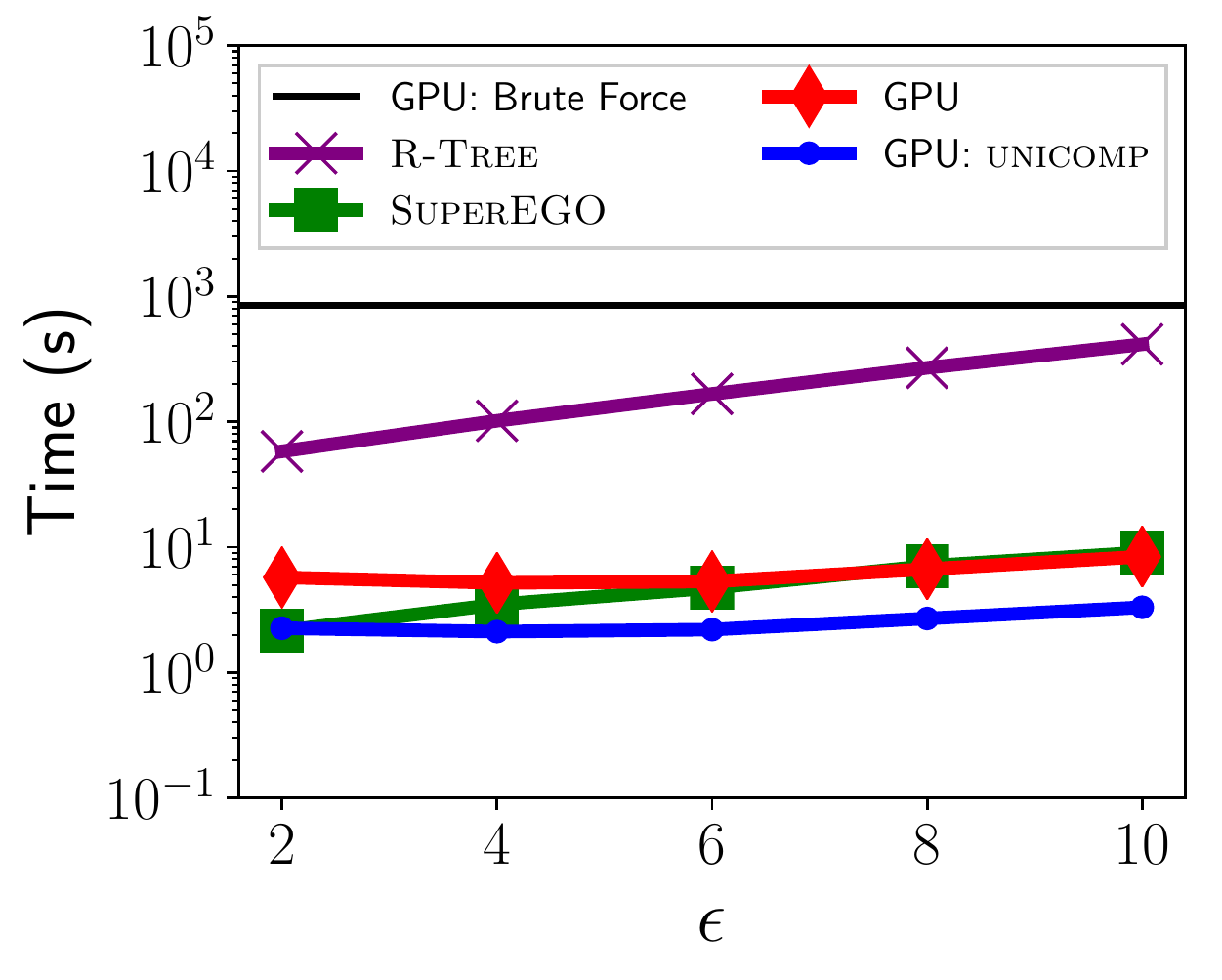}
    }
    \caption{Response time vs. $\epsilon$ on the 2--6 dimensional synthetic datasets with $2\times10^6$ points.}
   \label{fig:syn2M}
\end{figure*}

\begin{figure*}[!thp]
\centering
\subfigure[\dsynthba]{
        \includegraphics[width=0.181\textwidth, trim={0.7cm 0 0.7cm 0}]{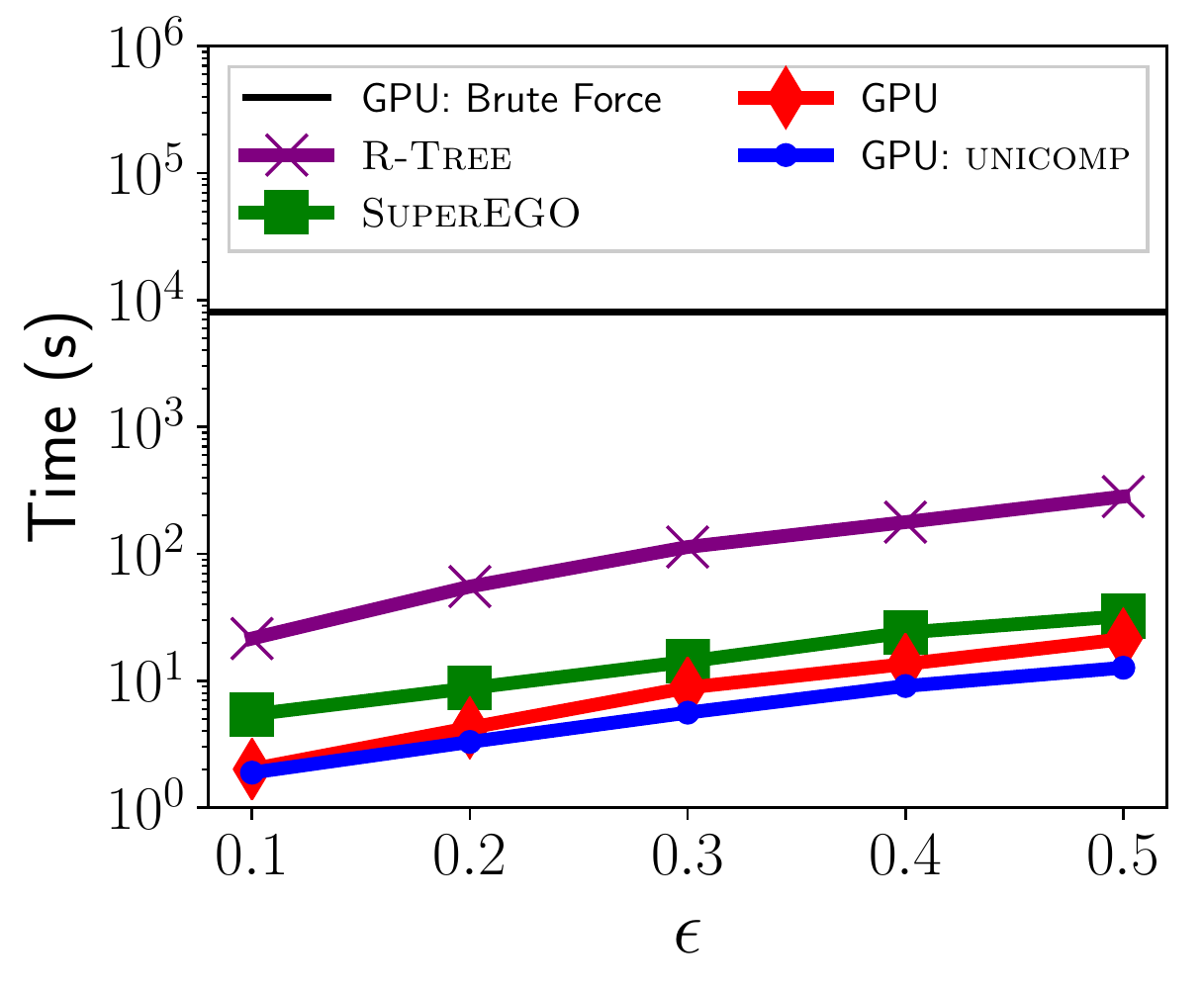}
    }
\subfigure[\dsynthbb]{
        \includegraphics[width=0.181\textwidth, trim={0.7cm 0 0.7cm 0}]{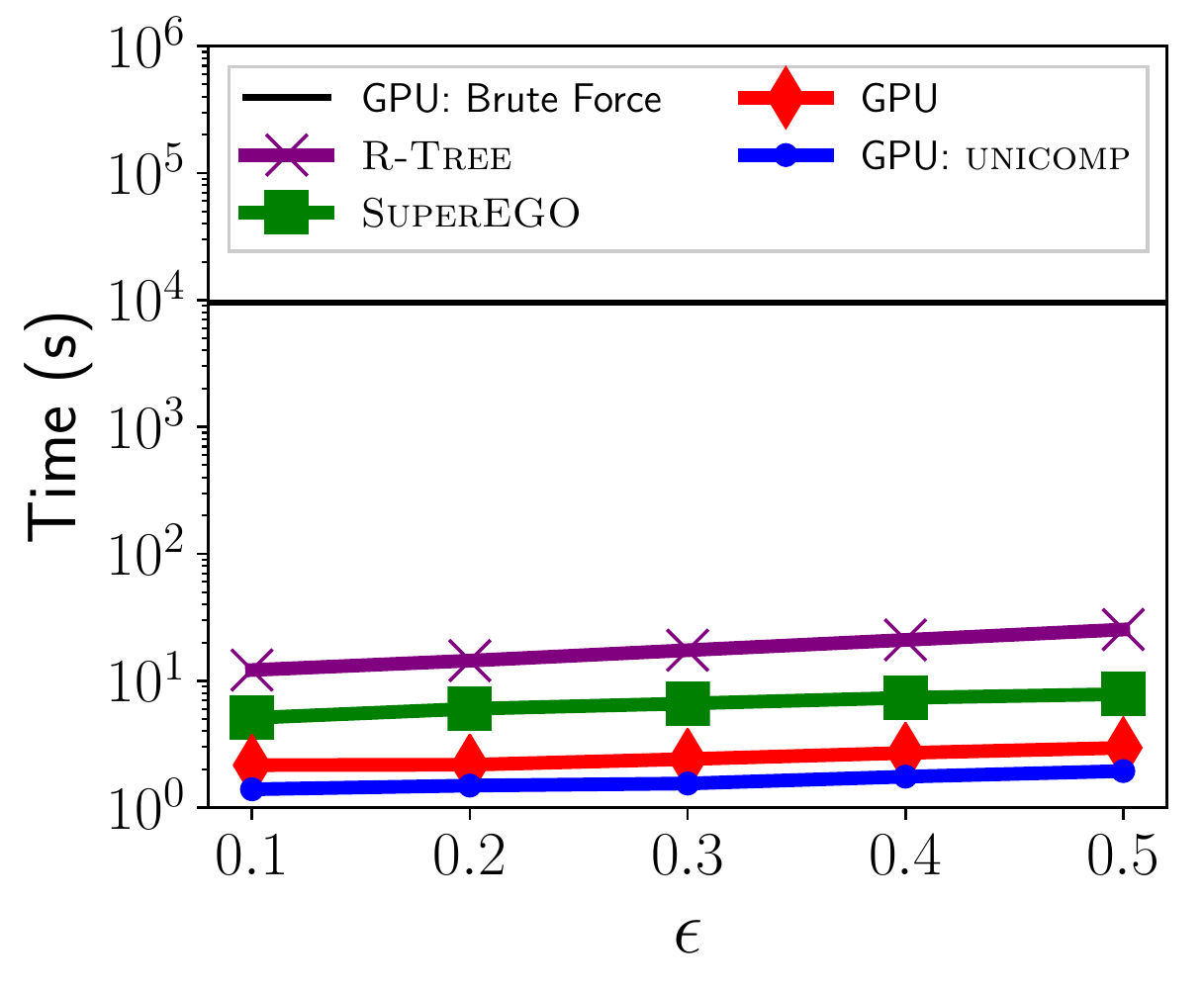}
    }
\subfigure[\dsynthbc]{
        \includegraphics[width=0.181\textwidth, trim={0.7cm 0 0.7cm 0}]{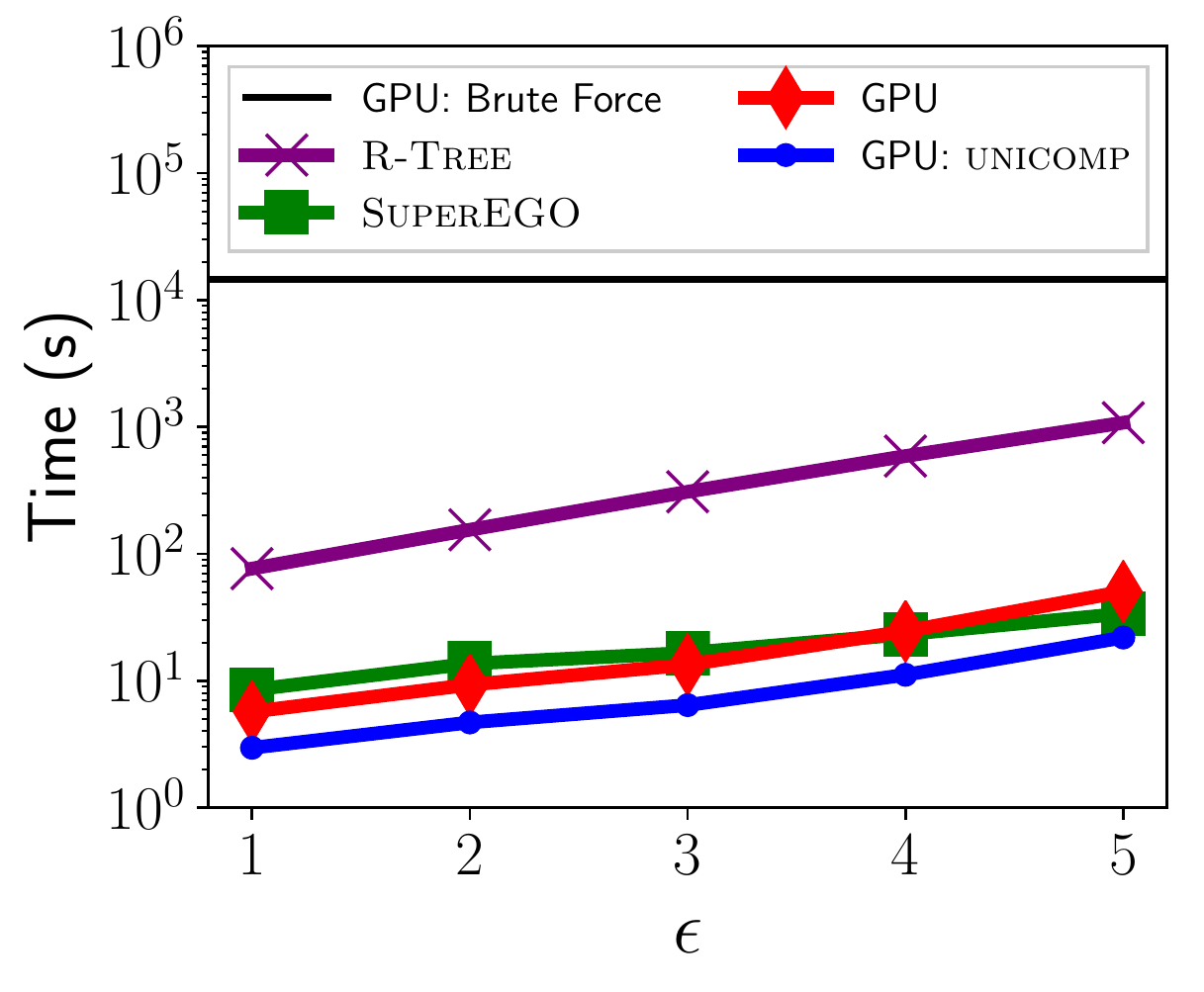}
    }
\subfigure[\dsynthbd]{
        \includegraphics[width=0.181\textwidth, trim={0.7cm 0 0.7cm 0}]{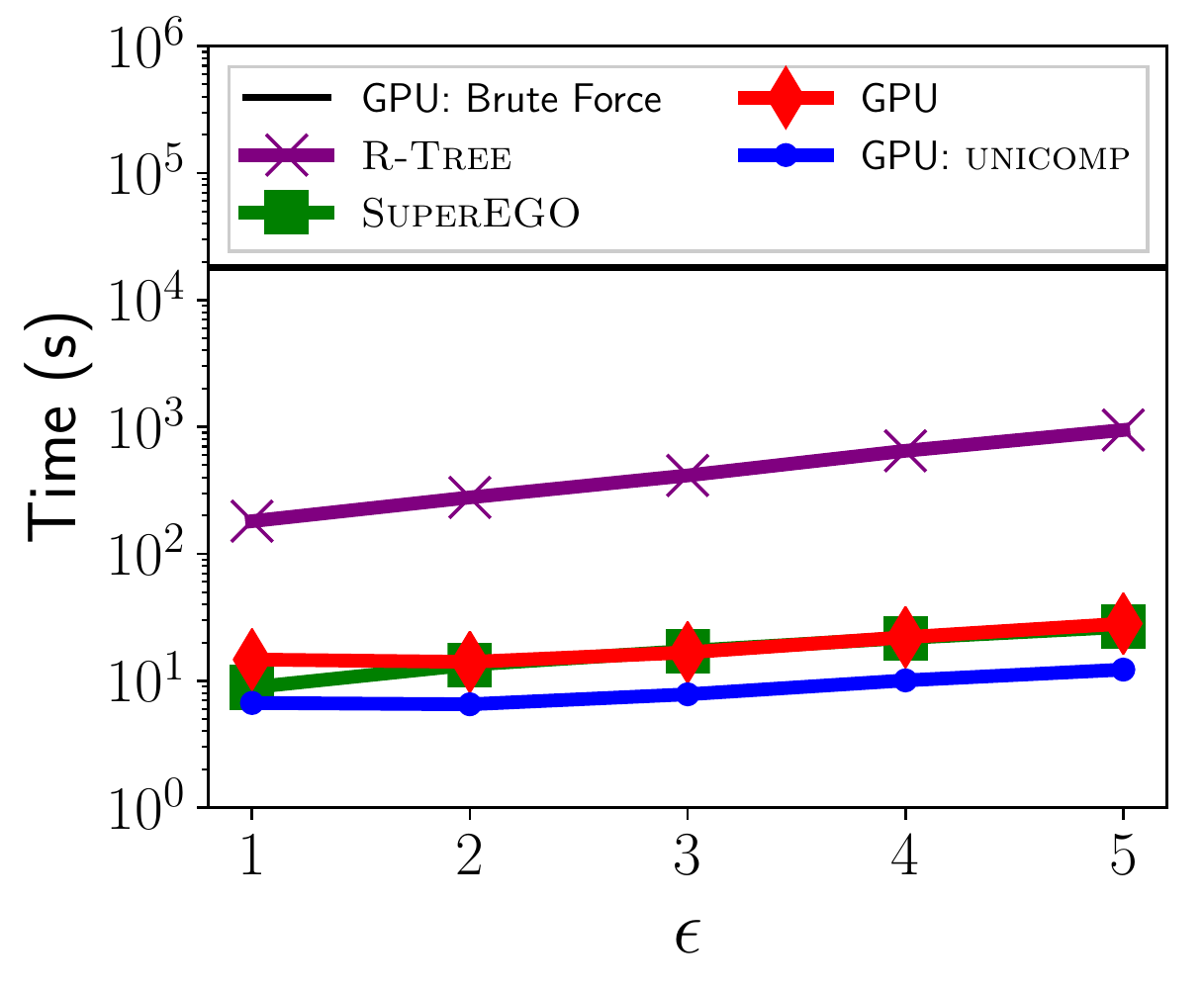}
    }
\subfigure[\dsynthbe]{
        \includegraphics[width=0.181\textwidth, trim={0.7cm 0 0.7cm 0}]{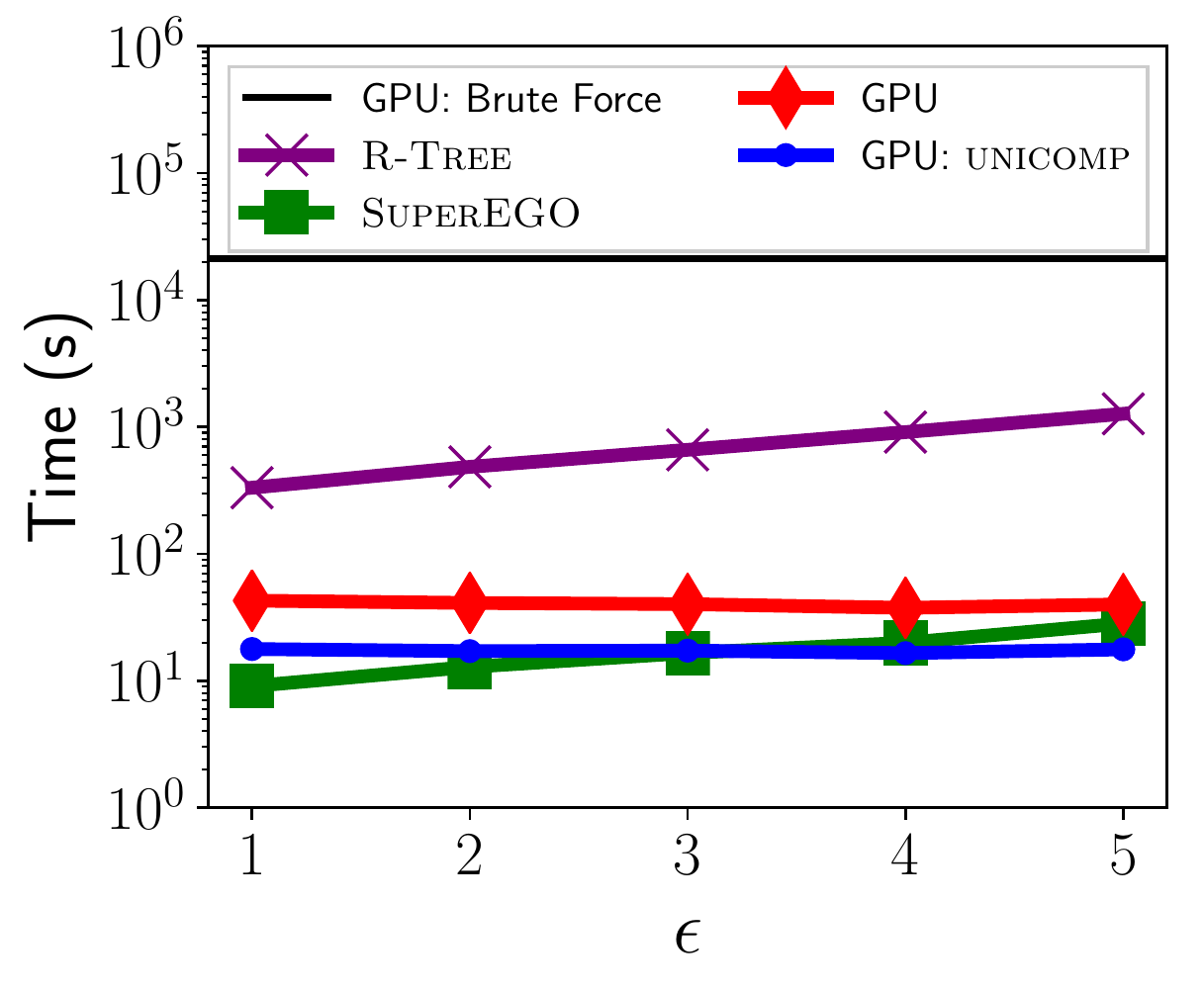}
    }
    \caption{Response time vs. $\epsilon$ on the 2--6 dimensional synthetic datasets with $10^7$ points.}
   \label{fig:syn10M}
\end{figure*}

\noindent \textbf{Synthetic datasets: } Figures~\ref{fig:syn2M}~and~\ref{fig:syn10M} show the response time vs. $\epsilon$ for synthetic datasets with 2 and 10 million datapoints, respectively, spanning 2--6 dimensions. While the results on the 2-D datasets are consistent with the observations for the real-world datasets in Figure~\ref{fig:real_world}, the results for higher dimensions are not. The performance of \unicomp improves with dimensionality, particularly when the number of dimensions, $n\geq3$ (comparing Figures~\ref{fig:syn2M}~and~\ref{fig:syn10M} (b)-(e)). Across all synthetic datasets, \gpu (with \unicomp) outperforms \rtree.  Furthermore, \gpu with \unicomp outperforms \ego on nearly all experiments.

\par
\textbf{Impact of data distribution on performance: }
We use uniformly distributed synthetic datasets because: $(i)$ the performance of searches is  data-dependent and uniformly distributed data is an average case compared to sparse datasets or datasets with over-dense regions; and $(ii)$ uniformly distributed data is a worst-case scenario for the \gpu grid index. Datasets having more over-dense regions will have fewer non-empty cells, resulting in fewer cells to search. In contrast, uniformly distributed data will have more cells with fewer points contained within, maximizing the number of non-empty cells. Thus, uniformly distributed data has greater \emph{search} overhead than data distributions with highly varying densities (i.e., real-world data).

\ego also tends to perform better on non-uniform data for many of the same reasons as \gpu. However, \ego reorders the dimensions to improve cell pruning as a function of data distribution~\cite{kalashnikov2013}. We find that across the uniformly distributed synthetic datasets, the performance of \ego degrades with $\epsilon$ in a similar manner to \gpu.

\begin{figure}[t]
\centering
        \includegraphics[width=0.4\textwidth, trim={0.7cm 0 0.7cm 0}]{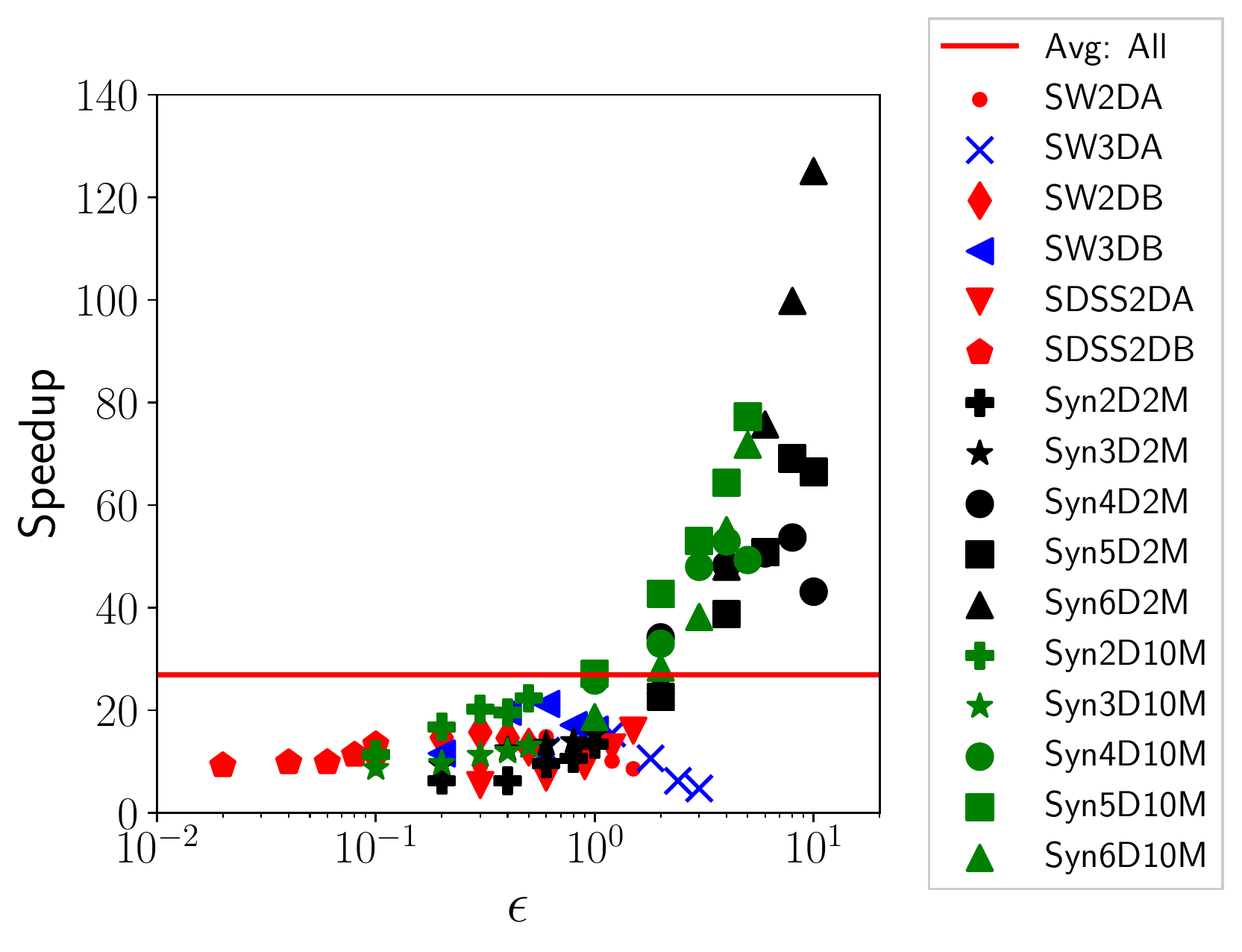}
    \caption{Speedup of \gpu with \unicomp over \rtree (1 thread) for all real-world and synthetic datasets plotted on log scale to capture orders of magnitude differences in $\epsilon$. Figure derived from Figures~\ref{fig:real_world},~\ref{fig:syn2M},~and~\ref{fig:syn10M}. The red line shows the average speedup (26.9$\times$).}
   \label{fig:comparison_real_syn_rtree}
\end{figure}

\textbf{\gpu vs. \rtree: }
Figure~\ref{fig:comparison_real_syn_rtree} plots the speedup of \gpu with \unicomp over \rtree vs. $\epsilon$ for each dataset from Figures~\ref{fig:real_world},~\ref{fig:syn2M},~and~\ref{fig:syn10M}.  The lowest performance gain over \rtree occurs on \sdssa and \dgeoaa (i.e., the smallest workloads).  When $2\leq n \leq 3$, we see fairly consistent speedups, regardless of $\epsilon$ or $|D|$.  This indicates that the speedup of \gpu over \rtree is less a function of data distribution (e.g., uniform vs. skewed).  Rather, data dimensionality dictates the rate at which $\epsilon$ degrades index performance, and this relationship differs between \gpu and \rtree.  This suggests that the speedup we see with
synthetic datasets when $4 \leq n \leq 6$ (Figure~\ref{fig:comparison_real_syn_rtree}) will be
consistent with real-world datasets of the same dimensionality.
Furthermore, index degradation with dimensionality explains why the performance gains are largest on the higher dimensional datasets (up to 125$\times$). Across all datasets, \gpu is on average 26.9$\times$ faster than \rtree.

\begin{figure}[t]
\centering
        \includegraphics[width=0.4\textwidth, trim={0.7cm 0 0.7cm 0}]{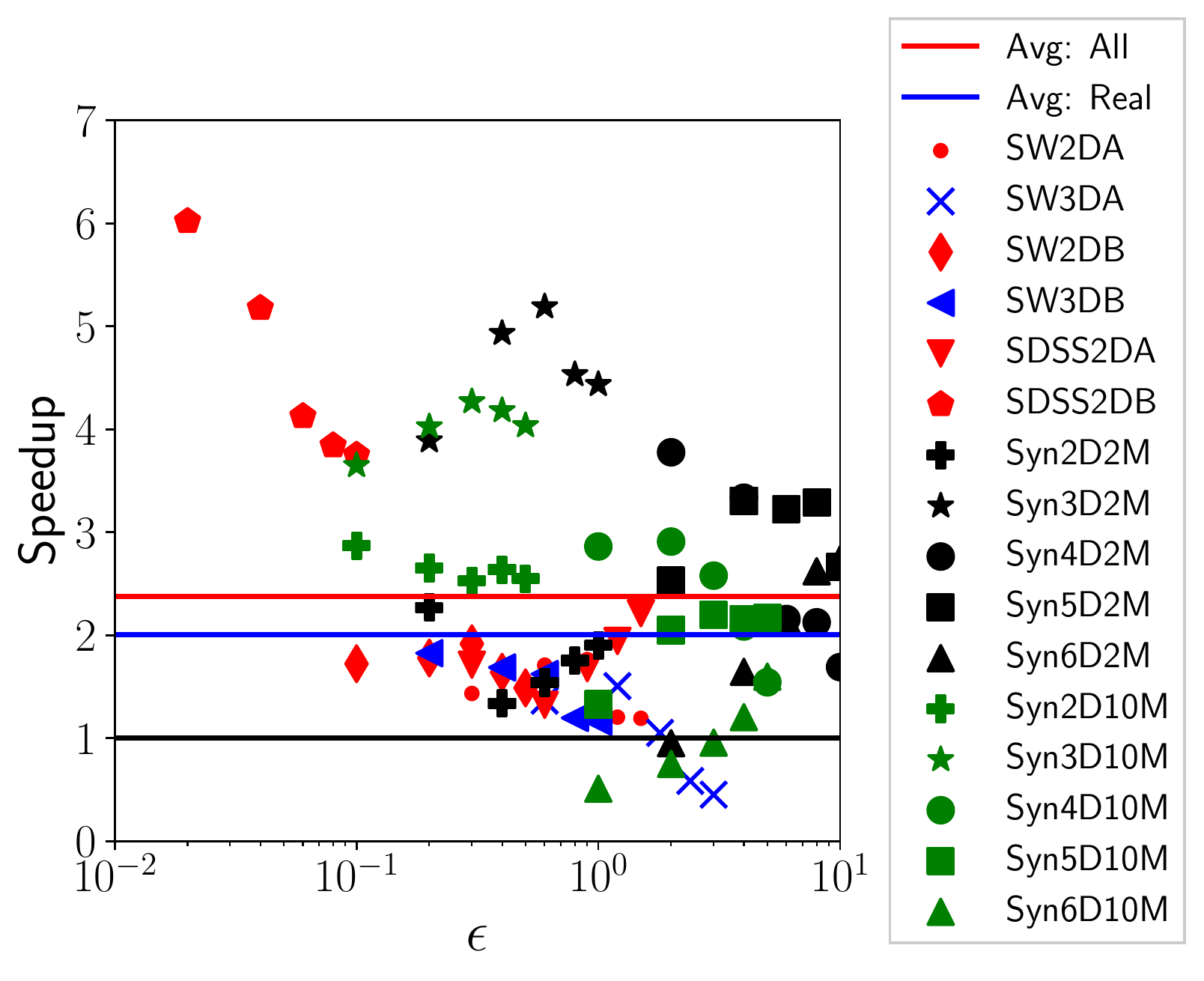}
    \caption{Speedup of \gpu with \unicomp over \ego (32 threads) for all real-world and synthetic datasets plotted on log scale to capture orders of magnitude differences in $\epsilon$. Figure derived from Figures~\ref{fig:real_world},~\ref{fig:syn2M},~and~\ref{fig:syn10M}. The black line shows where our approach achieves a speedup (or slowdown). The red  and blue lines show the average speedup across all of the datasets and real-world datasets, respectively.}
   \label{fig:comparison_2_3_D_real_syn_ego}
\end{figure}

\begin{figure*}[!htp]
\centering
\subfigure[Real: \datasetgeo ($2\leq n \leq 3$), \datasetsdss ($n=2$).]{
        \includegraphics[width=0.3\textwidth]{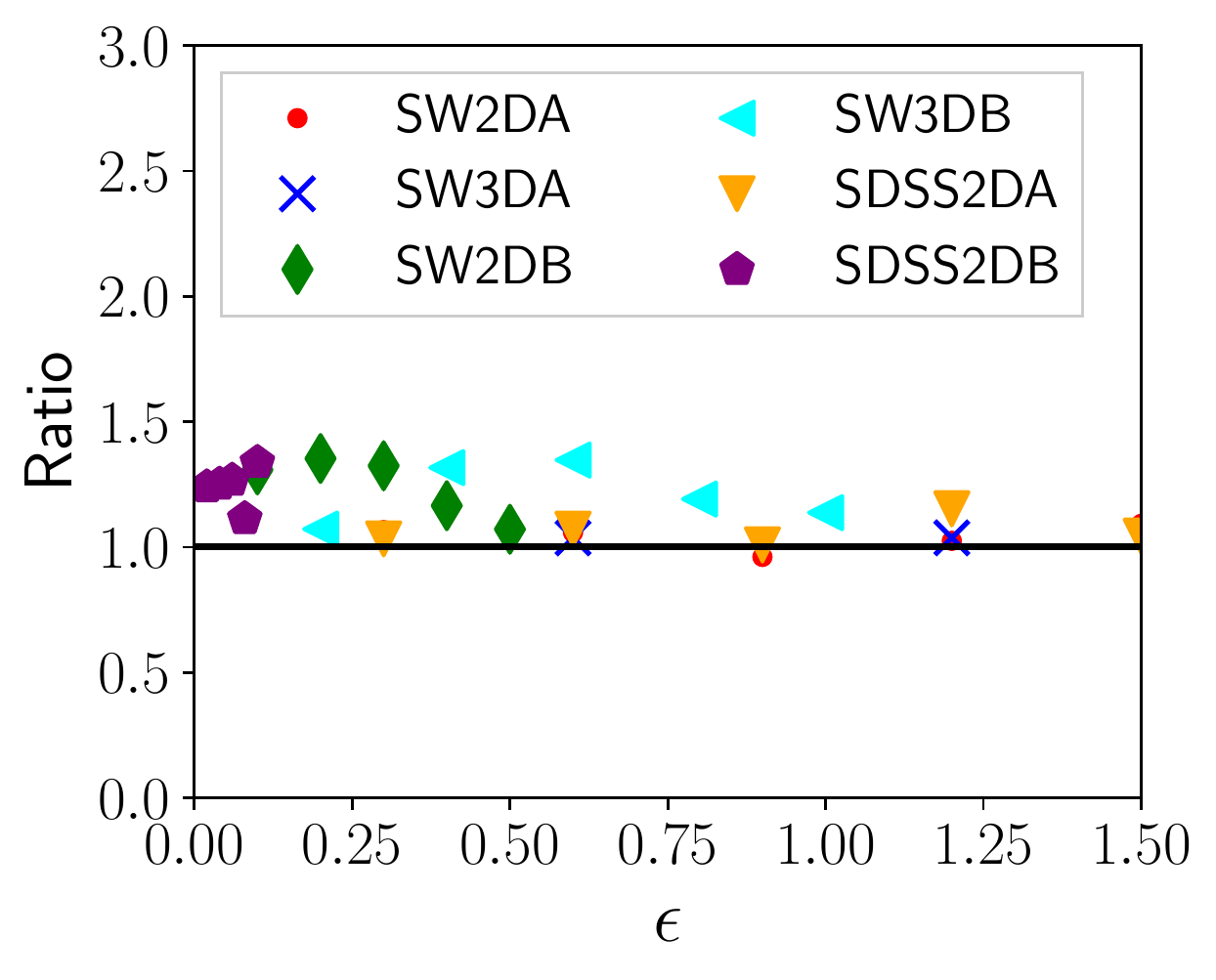}
    }
\subfigure[\datasetsyn $2\leq n \leq 6$, $|D|=2\times10^6$.]{
        \includegraphics[width=0.3\textwidth]{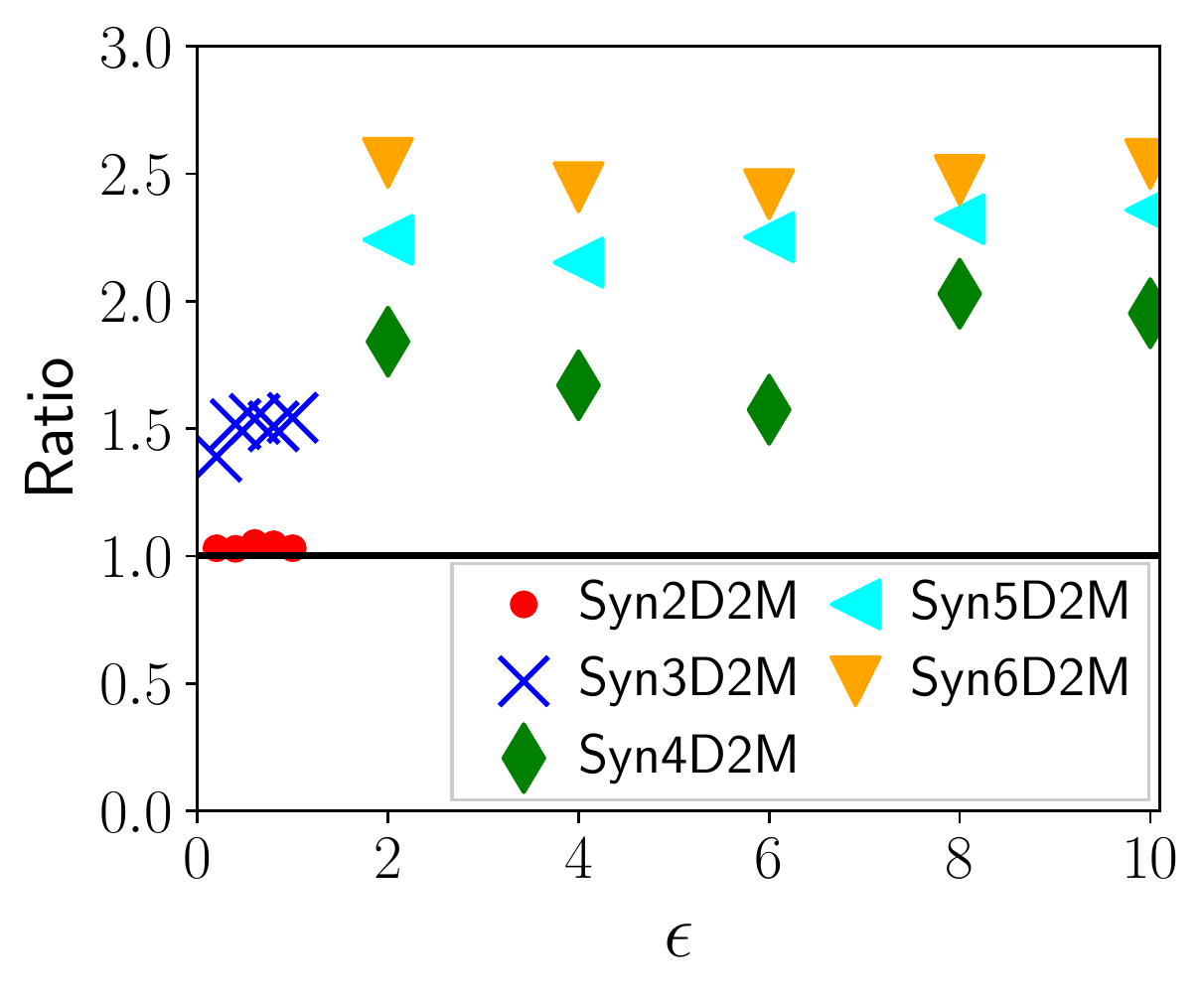}
    }
\subfigure[\datasetsyn $2\leq n \leq 6$, $|D|=10^7$.]{
        \includegraphics[width=0.3\textwidth]{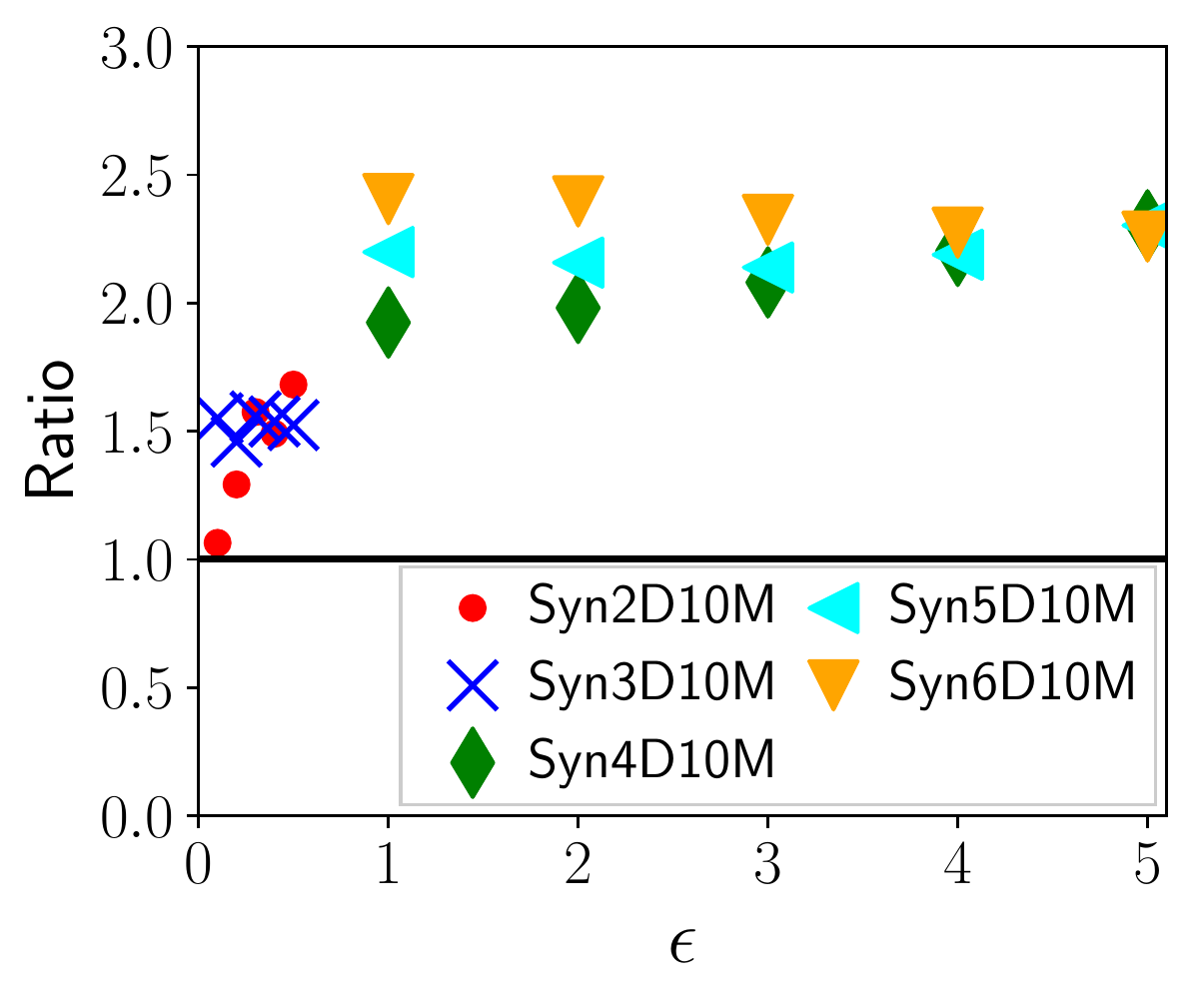}
    }
    \caption{Impact of the \unicomp optimization shown as the ratio of the response times of the GPU self-join without and with \unicomp. Panels (a), (b), and (c),  are derived from Figures~\ref{fig:real_world},~\ref{fig:syn2M}, and~\ref{fig:syn10M}, respectively. Points above the horizontal line indicate that \unicomp leads to a performance gain, whereas points below indicate a slowdown due to associated overheads.}
   \label{fig:ratio_unicomp}
\end{figure*}

\begin{table*}[htp]
\centering
\begin{footnotesize}
\caption{Selected kernel metrics of \gpu without and with \unicomp. All ratios given in the quantity of the metric with \unicomp divided by the value without the \unicomp optimization.}
\label{tab:compare_kernel_metrics}
\begin{tabular}{|c|c|c|c|c|c|c|c|c|} \hline
\makecell{Dataset} & \makecell{$\epsilon$}& \makecell{Ratio\\Resp. Time\\ (Figure~\ref{fig:ratio_unicomp})} & \makecell{Theoretical \\Occupancy\\GPU}&\makecell{Unified Cache \\Bandwidth\\ Utilization (GB/s)\\GPU}&\makecell{Theoretical \\Occupancy \\(GPU: \unicomp)}& \makecell{Unified Cache \\Bandwidth\\ Utilization (GB/s) \\(GPU: \unicomp)} & \makecell{Ratio\\Occupancy}& \makecell{Ratio\\Cache\\Utilization}\\\hline
\dgeoaa&0.3&1.07&100\%&597.87 &75\%&460.96 &0.75&0.77\\\hline
\sdssa&0.3&1.03&100\%&1157.07&75\%&864.26&0.75&0.75\\\hline
\dsynthad&8&\textbf{2.32}&62.5\%&306.45&50\%&578.14&0.8&\textbf{1.88}\\\hline
\dsynthae&8&\textbf{2.47}&62.5\%&293.31&50\%&469.24&0.8&\textbf{1.59}\\\hline
\end{tabular}
\end{footnotesize}
\end{table*}

\textbf{\gpu vs. \ego: }
Figure~\ref{fig:comparison_2_3_D_real_syn_ego} plots the speedup of \gpu with \unicomp over
\ego (with 32 threads).  
Similar to Figure~\ref{fig:comparison_real_syn_rtree}, the GPU performance gain is lowest on scenarios with small workloads. We find that there are only 6 instances where \ego outperforms \gpu (below the black line). The average speedup on real-world datasets is $\sim2\times$ (blue line), while the average across all datasets is $2.38\times$ (red line).  On average \gpu with \unicomp is 138\% faster than \ego. As expected, \ego performs worse on the synthetic datasets, as it cannot benefit from dimensionality reordering on uniformly distributed data.

Many-core GPUs can be leveraged in heterogeneous environments for data-intensive (self-) join operations. \ego is a state-of-the-art join algorithm, and across nearly all scenarios, our \gpu algorithm outperforms \ego as executed on 32 CPU cores. Also, recall that our results for \ego are using 32-bit data points, while \gpu uses 64-bit on all scenarios; therefore, we expect further performance gains over \ego if 64-bit floats are used.

\par
\textbf{Performance characterization of \unicomp: }
As previously mentioned, while \unicomp decreases the number of point comparisons and cells searched by a factor of $\sim$2, it does not yield a corresponding decrease in response time. This leads to unexpected speedups over \rtree and \ego as a function of $\epsilon$ and dataset size, as mentioned above.
Figure~\ref{fig:ratio_unicomp} shows the ratio of the response times of \gpu with and without \unicomp, derived from the results shown in Figures~\ref{fig:real_world},~\ref{fig:syn2M}, and~\ref{fig:syn10M} (by dividing the respective response times). While there are are a few scenarios in Figure~\ref{fig:ratio_unicomp}~(a) for which \unicomp results in a slight performance loss due to overhead, the resulting slowdowns are negligible; thus, \unicomp can be used without concern of significant performance degradation.

When using \unicomp on real-world datasets (Figure~\ref{fig:ratio_unicomp}~(a)), the response time ratios are within 1.5$\times$, and not 2$\times$ as expected. However, the higher dimensionality ($n\geq3$) results shown in Figure~\ref{fig:ratio_unicomp}~(b)~and~(c) demonstrate that \unicomp achieves some performance gains that are $\geq2\times$. This is a surprising result, as \unicomp reduces the number of cells and points searched by a factor of $\sim$2.
To understand this phenomenon, we consider two cases: when the response time ratio is $<2$
and when it is $>2$.  We use the nVIDIA Visual Profiler~\cite{NSIGHT} to collect two metrics during execution: \emph{occupancy} and \emph{unified cache utilization}.  Occupancy is a measure of the number of threads running simultaneously on the GPU.  Thus, low occupancy can result in under-utilization of GPU cores. On the nVIDIA Maxwell and Pascal GPUs, the unified (L1) cache is a coalescing buffer for memory accesses~\cite{PascalPerformance}. Particularly relevant is the caching of global loads.  We compare the occupancy and unified L1 cache bandwidth utilization on the self-join kernels (without and with \unicomp).

Table~\ref{tab:compare_kernel_metrics} shows two datasets with response time ratios of \gpu (without/with \unicomp) $<2$ (\dgeoaa, \sdssa) and $>2$ (\dsynthad, \dsynthae). In all datasets,
using \unicomp results in lower occupancy due to more registers being used per thread.  
As expected higher dimensionality also reduces occupancy due to register usage.  
Thus, while \unicomp reduces occupancy, we notice that the relative cache utilization depends
on the dataset.  For those with response time ratios $<2$ (\dgeoaa, \sdssa), \unicomp reduces
cache utilization.  However, when ratios are $>2$ (\dsynthad, \dsynthae), \unicomp \emph{increases}
cache utilization.  Thus, we attribute the increased performance of \unicomp on higher
dimensional datasets to a higher degree of temporal locality in the L1 cache.  This
explains the variance in performance from \unicomp, despite it decreasing the work 
by a factor of $\sim$2.

\section{Discussion \& Conclusion}\label{sec:conclusions}
The self-join is a widely used operation in many data intensive
search algorithms and Big Data applications. We demonstrate that our
algorithm, \gpu,  that combines a grid-based
index that is suited for the GPU with batched result
reporting and duplicate search removal (\unicomp), outperforms the multi-threaded 
state-of-the-art \ego algorithm with an average speedup of 2.38$\times$. \gpu also significantly outperforms the search-and-refine strategy, \rtree.

Due to the large range of relevant related work (Section~\ref{sec:background}), we summarize the novelty of our techniques as follows:

\noindent\textbullet In low-dimensionality, there are likely to be many neighbors within the query distance $\epsilon$. We batch the results to process self-join result sets that exceed the GPU's global memory capacity. The batching scheme overlaps computation and communication to amortize the host-GPU overheads.\\
\noindent\textbullet We motivate the use of a grid index for the GPU due to its bounded search, and thus  greater regular instruction flow in comparison to index-trees. While grids have been used on the CPU for the similarity join~\cite{bohm2001epsilon,kalashnikov2013}, we have not found any works that apply a grid on the GPU to compute the self-join.\\  
\noindent\textbullet By virtue of utilizing a grid, our novel \unicomp optimization reduces the number of grid cells searched and distance calculations, is generalizable to $n$-dimensions, and improves
cache utilization when $n>3$.

Future work includes applying
this work to other spatial searches, such as kNN, and determining if this work is relevant for self-joins on higher dimensional datasets.

\section*{Acknowledgment}
We thank Fr\'{e}d\'{e}ric Loulergue, and UHHPC at the University of Hawaii for the use of their  platforms.

\newcommand{\BIBdecl}{\setlength{\itemsep}{0.25 em}}
\bibliographystyle{IEEEtran}
{\tiny

}


\end{document}